\PassOptionsToPackage{hyphens}{url}
\PassOptionsToPackage{inline}{enumitem}
\def\isabridged{1}

\ifdefined\isabridged{}
\documentclass[letterpaper,twocolumn,10pt]{article}
\usepackage{usenix2019_v3}
\usepackage{amssymb,amsmath,amsthm}
\newtheorem{theorem}{Theorem}
\else
\documentclass[sigconf,nonacm]{acmart}
\fi

\usepackage[T1]{fontenc}
\usepackage[utf8]{inputenc}
\usepackage{letltxmacro}
\usepackage{todonotes}
\usepackage{booktabs}
\usepackage{amssymb}
\usepackage{amsmath}
\usepackage{xspace}
\usepackage{listings}
\usepackage{enumitem}
\usepackage{pifont}
\usepackage{hyperref}
\usepackage{multirow}
\usepackage{pbox}
\usepackage{color}
\usepackage[acronym]{glossaries}
\usepackage[olditem,oldenum]{paralist}
\usepackage{placeins}
\usepackage{subcaption}
\usepackage{adjustbox}
\usepackage{cleveref}
\usepackage[multiple]{footmisc}

\usepackage[lambda,operators,sets]{cryptocode}
\renewcommand{\sample}{\stackrel{\$}{\leftarrow}}
\newif\ifabridged{}
\ifdefined\isabridged{}\abridgedtrue{}\fi
\newif\ifnotabridged{}
\ifabridged\notabridgedfalse{}\else\notabridgedtrue{}\fi

\newif\ifanonymous{}
\ifdefined\isanonymous{}\anonymoustrue{}\fi
\newif\ifnotanonymous{}
\ifanonymous\notanonymousfalse{}\else\notanonymoustrue{}\fi

\newif\ifbw{}
\ifdefined\isbw{}\bwtrue{}\fi

\newif\ifaebadge{}
\ifdefined\isifaebadge{}\badgetrue{}\fi

\ifanonymous
\usepackage{titling}
\pretitle{\begin{center}\LARGE}
\posttitle{\par\end{center}}
\preauthor{}
\postauthor{}
\predate{}
\date{}
\postdate{}
\fi
\glsdisablehyper{}
\definecolor{lst:comment}{rgb}{0.0,0.0,0.0}
\definecolor{lst:instructions}{rgb}{0.4,0.1,0.1}
\definecolor{lst:registers}{rgb}{0.1,0.4,0.1}
\definecolor{lst:bg}{rgb}{0.98,0.98,0.98}
\lstdefinelanguage{aarch64}{
    keywords={pacga,stp,mov,bl,ldp,cmp,jnz,ret,str,ldr,autia,pacia,eor,autib,pacib,b,cbnz,cbz,paciasp,retaa},
    keywordstyle=\bfseries\color{lst:instructions},
    keywords=[2]{LR, FP, CR, Xd, Xr, SP,XZR, X28, X15},
    keywordstyle=[2]\color{lst:registers},
    breaklines=true,
    morecomment=[l]{;},
    commentstyle=\itshape\color{lst:comment},
}
\lstdefinestyle{customasm}{language=aarch64}
\lstdefinestyle{customc}{language=c,
    keywordstyle=\bfseries\color{green!40!black},
    commentstyle=\itshape\color{purple!40!black},
    identifierstyle=\color{blue},
    stringstyle=\color{orange},
}

\ifabridged
\newcommand{\inputCListing}[2]{\lstinputlisting[basicstyle=\ttfamily\footnotesize,belowskip=-0.8 \baselineskip,float=,floatplacement=tp,showstringspaces=false,style=customc,linerange={3-100},label={lst:#1},caption={#2}]{listings/#1.c}}

\newcommand{\inputAsmListing}[2]{\lstinputlisting[basicstyle=\ttfamily\footnotesize,belowskip=-0.8 \baselineskip,float=,floatplacement=tp,showstringspaces=false,style=customasm,linerange={2-100},label={lst:#1},caption={#2}]{listings/#1.s}}

\else
\newcommand{\inputCListing}[2]{\lstinputlisting[float=,floatplacement=tp,showstringspaces=false,style=customc,linerange={3-100},label={lst:#1},caption={#2}]{listings/#1.c}}

\newcommand{\inputAsmListing}[2]{\lstinputlisting[float=,floatplacement=tp,showstringspaces=false,style=customasm,linerange={2-100},label={lst:#1},caption={#2}]{listings/#1.s}}

\fi
\LetLtxMacro{\todonote}{\todo}
\renewcommand{\todo}[2][]
{\todonote[inline, caption={#2}, size=\footnotesize, #1]
{\renewcommand{\baselinestretch}{0.5}\selectfont#2\par}}

\newcommand{\SHORTNAME}{\protect{ACS}\xspace}
\newcommand{\ARMPA}{\protect{\gls{pa}}\xspace}
\DeclareRobustCommand{\ARMPAFULL}{\protect{ARMv8.3-A \gls{pa}}\xspace}

\newcommand{\LONGNAME}{\protect{authenticated call stack}\xspace}
\newcommand{\IMPLNAME}{\protect{\textsf{PACStack}}\xspace}

\newcommand{\nginx}{\protect{NGINX}\xspace}

\newcommand{\overhead}{$\approx$3\%\xspace}
\newcommand{\overheadNbench}{0.5\%\xspace}

\newcommand{\overheadSCpp}{0.9\%\xspace}
\newcommand{\overheadSCppNomask}{2.0\%\xspace}

\newcommand{\etal}{et al.\xspace}

\newcommand{\shadowstack}{\textsf{ShadowCallStack}\xspace}
\newcommand{\nomaskpacstack}{\textsf{\IMPLNAME-nomask}\xspace}

\newcommand{\version}[1]{\ensuremath{\mathrm{v}#1}\xspace}

\newcommand{\material}{\url{https://drive.google.com/drive/folders/1IbAfv7ccViZTcX8BC1_ltogMHi-bmYUU}}

\newcommand{\instr}[1]{\texttt{\lowercase{#1}}\xspace}
\newcommand{\func}[1]{\texttt{#1}\xspace}
\newcommand{\setjmp}{\func{setjmp}}
\newcommand{\longjmp}{\func{longjmp}}
\newcommand{\setjmpWrapper}{\func{setjmp\_wrapper}}
\newcommand{\longjmpWrapper}{\func{longjmp\_wrapper}}
\newcommand{\jmpbuf}{\texttt{jmp\_buf}\xspace}
\newcommand{\pachash}[2]{\ensuremath{\mathtt{H_K}(#1,#2)}\xspace}
\newcommand{\pacmask}[1]{\ensuremath{\mathtt{H_K}(0, #1)}\xspace}
\newcommand{\pactag}[2]{\ensuremath{\mathtt{T_K}(#1,#2)}\xspace}
\newcommand{\pacg}[2]{\ensuremath{\mathtt{pacga}(#1,#2)}\xspace}
\newcommand{\paci}[2]{\ensuremath{\mathtt{pacia}(#1,#2)}\xspace}
\newcommand{\auti}[2]{\ensuremath{\mathtt{autia}(#1,#2)}\xspace}

\newcommand{\pacgplain}{\ensuremath{\mathtt{pacga}}\xspace}
\newcommand{\paciplain}{\ensuremath{\mathtt{pacia}}\xspace}
\newcommand{\autiplain}{\ensuremath{\mathtt{autia}}\xspace}
\newcommand{\mReg}[1]{\ensuremath{\mathtt{#1}}}

\newcommand{\auth}[1]{\ensuremath{auth_{#1}}}
\newcommand{\authAdv}[1]{\ensuremath{auth'_{#1}}}

\newcommand{\ret}[1]{\ensuremath{ret_{#1}}}
\newcommand{\retAdv}[1]{\ensuremath{ret'_{#1}}}
\newcommand{\retBad}[1]{\ensuremath{ret^*_{#1}}}
\newcommand{\aret}[1]{\ensuremath{aret_{#1}}}
\newcommand{\aretAdv}[1]{\ensuremath{aret'_{#1}}}

\newcommand{\sigret}[1]{\ensuremath{sigret_{#1}}}
\newcommand{\sigretAdv}[1]{\ensuremath{sigret'_{#1}}}
\newcommand{\asigret}[1]{\ensuremath{asigret_{#1}}}
\newcommand{\asigretAdv}[1]{\ensuremath{asigret'_{#1}}}

\newcommand{\crDef}{\texttt{X28}\xspace}
\newcommand{\CR}{\gls{cr}\xspace}

\newcommand{\libc}{\texttt{libc}\xspace}
\newcommand{\libunwind}{\texttt{libunwind}\xspace}

\newcommand{\theAttacker}{\ensuremath{\mathcal{A}}\xspace}

\newcommand{\rowPolicy}{W$\oplus{}$X\xspace}

\newcommand{\branchProtFlag}{\texttt{-mbranch-protection}\xspace}
\newcommand{\stackProtFlag}{\texttt{-mstack-protector-strong}\xspace}
\newcommand{\dOne}{\ding{182}\xspace}\newcommand{\dTwo}{\ding{183}\xspace}\newcommand{\dThree}{\ding{184}\xspace}
\newcommand{\dFour}{\ding{185}\xspace}\newcommand{\dFive}{\ding{186}\xspace}\newcommand{\dSix}{\ding{187}\xspace}
\newcommand{\dSeven}{\ding{188}\xspace}
\newcommand{\dCOne}{\ding{192}\xspace}\newcommand{\dCTwo}{\ding{193}\xspace}\newcommand{\dCThree}{\ding{194}\xspace}
\newcommand{\dCFour}{\ding{195}\xspace}\newcommand{\dCFive}{\ding{196}\xspace}\newcommand{\dCSix}{\ding{197}\xspace}

\newcommand{\acsFastProStrAut}{\dCOne}
\newcommand{\acsFastProStrFrameRecord}{\dCTwo}
\newcommand{\acsFastProPac}{\dCThree}
\newcommand{\acsFastProCR}{\dCFour}
\newcommand{\acsFastEpiLdrAut}{\dCFive}
\newcommand{\acsFastEpiAut}{\dCSix}
\newcommand{\acsMaskProPacMask}{\dOne}
\newcommand{\acsMaskProEorMask}{\dTwo}
\newcommand{\acsMaskProMaskClear}{\dThree}
\newcommand{\acsMaskEpiLdrAut}{\dFour}
\newcommand{\acsMaskEpiPacMask}{\dFive}
\newcommand{\acsMaskEpiEorMask}{\dSix}
\newcommand{\acsMaskEpiMaskClear}{\dSeven}

\newcommand{\tableNo}{\ding{55}}
\newcommand{\tableYes}{\ding{51}}

\newcommand{\adv}{\textsf{Adv}}

\graphicspath{{figures/}}
\setlist[enumerate]{topsep=0pt,itemsep=-1ex,partopsep=1ex,parsep=1ex}
\setlist[itemize]{topsep=0pt,itemsep=-1ex,partopsep=1ex,parsep=1ex}
\makeatletter
\def\blfootnote{\xdef\@thefnmark{}\@footnotetext}
\makeatother
\newacronym{aslr}{ASLR}{address space layout randomization}
\newacronym{bti}{BTI}{Branch Target Indicator}
\newacronym{cfg}{CFG}{control-flow graph}
\newacronym{cfi}{CFI}{control-flow integrity}
\newacronym{cps}{CPS}{connections per second}
\newacronym{cr}{\mReg{CR}}{chain register}
\newacronym{dep}{DEP}{data-execution prevention}
\newacronym{dop}{DOP}{data-oriented programming}
\newacronym{el}{EL}{exception level}
\newacronym{fvp}{FVP}{Fixed Virtual Platform}
\newacronym{ir}{IR}{Intermediate Representation}
\newacronym{lr}{\mReg{LR}}{link register}
\newacronym{mac}{MAC}{message authentication code}
\newacronym{mmu}{MMU}{memory management unit}
\newacronym{pac}{PAC}{pointer authentication code}
\newacronym{pa}{PA}{pointer authentication}
\newacronym{pc}{\mReg{PC}}{program counter}
\newacronym{rop}{ROP}{return-oriented programming}
\newacronym{sp}{\mReg{SP}}{stack pointer}
\newacronym{ssltps}{SSL TPS}{SSL/TLS transactions per second}
\newacronym{toctou}{TOCTOU}{time-of-check-time-of-use}
\newacronym{va}{VA}{virtual address}
\newacronym{vdso}{\texttt{vdso}}{virtual dynamic shared object}
\newacronym[longplural=systems-on-chip]{soc}{SoC}{system on chip}
\ifaebadge{}
\usepackage[firstpage]{draftwatermark}
\SetWatermarkText{\hspace*{6in}\raisebox{8.5in}{\includegraphics{figures/usenix_artifact_evaluation_passed_new.pdf}}}
\SetWatermarkAngle{0}
\fi

\begin{document}

\title{PACStack: an Authenticated Call Stack}

\ifnotanonymous{}
\ifabridged{}
\author{
    {\rm Hans Liljestrand}\\
    University of Waterloo, Canada\\
    {\rm hans@liljestrand.dev}
    \and
    {\rm Thomas Nyman}\\
    Aalto University, Finland\\
    {\rm thomas.nyman@aalto.fi}
    \and
    {\rm Lachlan J.\ Gunn}\\
    Aalto University, Finland\\
    {\rm lachlan@gunn.ee}
    \and
    {\rm Jan-Erik Ekberg}\\
    Huawei Technologies Oy, Finland\\
    Aalto University, Finland\\
    {\rm jan.erik.ekberg@huawei.com}
    \and
    {\rm N. Asokan}\\
    University of Waterloo, Canada\\
    Aalto University, Finland\\
    {\rm asokan@acm.org}
}
\else
\author{Hans Liljestrand}
\affiliation{\institution{University of Waterloo, Canada}}
\email{hans@liljestrand.dev}

\author{Thomas Nyman}
\affiliation{\institution{Aalto University, Finland}}
\email{thomas.nyman@aalto.fi}

\author{Lachlan J.\ Gunn}
\affiliation{\institution{Aalto University, Finland}}
\email{lachlan@gunn.ee}

\author{Jan-Erik Ekberg}
\affiliation{\institution{Huawei Technologies Oy, Finland}\institution{Aalto University, Finland}}
\email{jan.erik.ekberg@huawei.com}

\author{N. Asokan}
\affiliation{\institution{University of Waterloo, Canada}\institution{Aalto University, Finland}}
\email{asokan@acm.org}
\renewcommand{\shortauthors}{Liljestrand, et al.}
\renewcommand\footnotetextcopyrightpermission[1]{}
\settopmatter{printacmref=false,printfolios=true}
\fi
\pagenumbering{gobble}
\fi

\ifnotabridged{}
\begin{abstract}
A popular run-time attack technique is to
compromise the \glsdesc{cfi} of a program by modifying function return addresses on the stack.
So far, shadow stacks have proven to be essential for \emph{comprehensively preventing} return address manipulation.
Shadow stacks record return addresses in integrity-protected memory secured with hardware-assistance or software access control.
Software shadow stacks incur high overheads or trade off security for efficiency.
Hardware-assisted shadow stacks are efficient and secure, but require the deployment of special-purpose hardware.

We present \emph{\LONGNAME} (\SHORTNAME), an approach that uses chained \glspl{mac}.
Our prototype, \IMPLNAME, uses the ARM general purpose hardware mechanism for \gls{pa} to implement \SHORTNAME.
Via a rigorous security analysis, we show that \IMPLNAME achieves security comparable to hardware-assisted shadow stacks \emph{without requiring dedicated hardware}.
We demonstrate that \IMPLNAME's performance overhead is small (\overhead).
\end{abstract}

\fi

\maketitle

\ifanonymous{}
{\let\thefootnote\relax\footnote{A preliminary version of this was presented as a poster at an ACM conference (see anonymized supplementary material at \material)}}
\fi

\ifabridged{} 
\vspace*{-2.5em}
\begin{abstract}
A popular run-time attack technique is to
compromise the \glsdesc{cfi} of a program by modifying function return addresses on the stack.
So far, shadow stacks have proven to be essential for \emph{comprehensively preventing} return address manipulation.
Shadow stacks record return addresses in integrity-protected memory secured with hardware-assistance or software access control.
Software shadow stacks incur high overheads or trade off security for efficiency.
Hardware-assisted shadow stacks are efficient and secure, but require the deployment of special-purpose hardware.

We present \emph{\LONGNAME} (\SHORTNAME), an approach that uses chained \glspl{mac}.
Our prototype, \IMPLNAME, uses the ARM general purpose hardware mechanism for \gls{pa} to implement \SHORTNAME.
Via a rigorous security analysis, we show that \IMPLNAME achieves security comparable to hardware-assisted shadow stacks \emph{without requiring dedicated hardware}.
We demonstrate that \IMPLNAME's performance overhead is small (\overhead).
\end{abstract}

\fi 

\glsresetall
\section{Introduction}
\label{sec:introduction}

Traditional code-injection attacks are ineffective in the presence of \rowPolicy{} policies that prevent the modification of executable memory~\cite{Szekeres2013}.
However, code-reuse attacks can alter the run-time behavior of a program without
modifying any of its executable code sections.
\Gls{rop} is a prevalent attack technique that corrupts function return
addresses to hijack a program's control flow. \Gls{rop} can be used to achieve Turing-complete computation by chaining together existing code sequences in the victim program.
To prevent \gls{rop}, return addresses must be protected when stored in memory.
At present, the most powerful protection against \gls{rop} is using an \emph{integrity-protected shadow stack} that maintains a secure reference copy of each return address~\cite{Abadi09}.
Integrity of the shadow stack is ensured by making it inaccessible to the
adversary either by randomizing its location in memory or by using specialized
hardware~\cite{Intel-CET}.
Recent software-based shadow stacks show reasonable performance~\cite{Burow19},
but are vulnerable to an adversary capable of exploiting memory vulnerabilities
to infer the location of the shadow stack.
To date, only hardware-assisted schemes, such as Intel CET~\cite{Intel-CET}, achieve negligible overhead without trading off security.
But employing such a custom hardware mechanism incurs development and deployment costs.

Recent ARM processors include support for \ifnotabridged general-purpose\fi \gls{pa}; a hardware extension that uses tweakable \glspl{mac} to sign and verify pointers~\cite{ARMv8A}.
One initial use case of \gls{pa} is the authentication of return addresses~\cite{Qualcomm17}.
However, current \gls{pa} schemes are vulnerable to \emph{reuse attacks}, where the adversary can reuse previously observed valid protected pointers~\cite{Liljestrand19}.
Prior work~\cite{Qualcomm17,Liljestrand19} and current implementations by GCC
\footnote{\url{https://gcc.gnu.org/onlinedocs/gcc/AArch64-Function-Attributes.html}}
and LLVM
\footnote{\url{https://reviews.llvm.org/D49793}}
mitigate reuse attacks, but cannot completely prevent them.

In this paper, we propose a new approach, \emph{\LONGNAME} (\SHORTNAME), providing security comparable to hardware-assisted shadow stacks, with minimal overhead and without requiring new hardware-protected memory.
\SHORTNAME binds all return addresses into a chain of \glspl{mac} that allow verification of return addresses before their use.
We show how \SHORTNAME{} can be efficiently realized using ARM \gls{pa} while resisting reuse attacks.
The resulting system, \IMPLNAME{}, can withstand strong adversaries with full memory access.
Our contributions are:
\begin{itemize}[leftmargin=*]
\item \SHORTNAME{}, a new approach for \textbf{precise verification of function return addresses} by chaining \glspl{mac} (Section~\ref{sec:design}).
\item \IMPLNAME, an LLVM-based realization of \SHORTNAME{} using ARM \gls{pa} \textbf{without requiring additional hardware} (Section~\ref{sec:implementation}).
\item A systematic evaluation of \IMPLNAME security, showing that its \textbf{security is comparable to shadow stacks} (Section~\ref{sec:evaluation-security}).
\item Demonstrating that the \textbf{performance overhead} of \IMPLNAME \textbf{is small} (\overhead) (Section~\ref{sec:evaluation-performance}).
\end{itemize}
\ifnotabridged
For realizing \IMPLNAME, we implemented an efficient authenticated stack using ARM \gls{pa}\@. This approach may be generalizable to other data structures and applications (Section~\ref{sec:discussion-generalizing-acs}).
\fi
\IMPLNAME and associated evaluation code is available as open source \ifnotanonymous at \url{https://pacstack.github.io}\fi.

\section{Background}
\label{sec:background}

\subsection{ROP on ARM}
\label{sec:background-rop-arm}\label{sec:background-attacks}
In \gls{rop}, the adversary exploits a memory vulnerability to manipulate return addresses stored on the stack, thereby altering the program's backward-edge control flow.
\gls{rop} allows Turing-complete attacks by chaining together multiple gadgets, i.e., adversary-chosen sequences of pre-existing program instructions that together perform the desired operations.
ARM architectures use the \gls{lr} to hold the current function's return address.
\gls{lr} is automatically set by the \emph{branch with link} (\instr{bl}) or \emph{branch with link to register} (\instr{blr}) instructions that are used to implement regular and indirect function calls.
Because \gls{lr} is overwritten on call, non-leaf functions must store the return address onto the stack.
This opens up the possibility of \gls{rop} on ARM \ifnotabridged architectures\fi~\cite{Kornau2009}.

\subsection{ARM Pointer Authentication}
\label{sec:background-pa}

The \ARMPAFULL extension supports calculating and verifying \glspl{pac}~\cite{ARMv8A}.
\Gls{pa} is at present deployed in the Apple A12, A13, S4, and S5 \glspl{soc} and is going to be available in all upcoming ARMv8.3-A and later \glspl{soc}.
A \instr{pac} instruction calculates a keyed tweakable \gls{mac}, $\pachash{A_P}{M}$, over the address $A_P$ of a pointer $P$ using a 64-bit modifier $M$ as the tweak.
The resulting authentication token, referred to as a \gls{pac}, is embedded into the unused
high-order bits of $P$.
It can be verified using an \instr{aut} instruction that recalculates $\pachash{A_P}{M}$, and compares the result to $P$'s \gls{pac}\@.

\begin{figure}[tp]
\centering
  \ifbw\includegraphics[width=1\columnwidth]{pac-bw}
  \else\includegraphics[width=1\columnwidth]{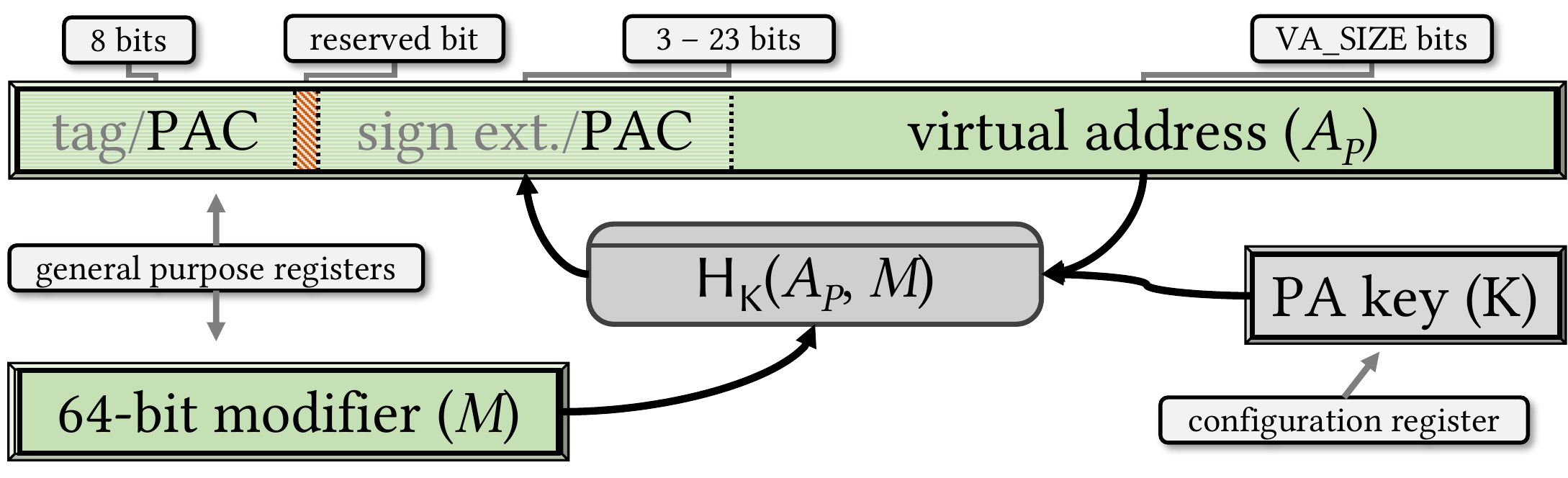}\fi
\caption{
\Gls{pa} uses a \glsfirst{pac} based on the pointer's address, a modifier, and a key.
}
\label{fig:pac}
\end{figure}

Since the \gls{pac} is stored in unused bits of a pointer, its size is limited by the virtual address size (\texttt{VA\_SIZE} in Figure~\ref{fig:pac}) and whether address tagging is enabled~\cite{ARMv8A}.
On a 64-bit ARM machine running a default Linux kernel, \texttt{VA\_SIZE} is 39, which leaves 16 bits for the \gls{pac} when excluding the reserved and address tag bits.
\gls{pa} provides five different keys; two for code pointers, two for data pointers, and one for generic use.
Each key has a separate set of instructions\ifnotabridged\footnote{A full list of \gls{pa} instructions from~\cite{Liljestrand19} is available in Appendix~\ref{sec:appendix-instructions}.}\fi, e.g., the \instr{autia} and \instr{pacia} instructions always operate on the instruction key $A$, stored in the \texttt{APIAKey\_EL1} register.
Access to the key registers and \gls{pa} configuration registers can be restricted to a higher \gls{el}\@.
Linux \version{5.0}
\footnote{\url{https://kernelnewbies.org/Linux_5.0\#ARM_pointer_authentication}}
adds full support for \gls{pa}, such that the kernel (at \gls{el}1) manages user-space (\gls{el}0) keys and prevents \gls{el}0 from modifying them.
The kernel generates new \gls{pa} keys for a process on an \texttt{exec} system call.

As currently specified, \gls{pa} does not cause a fault on verification failure; instead, it strips the \gls{pac} from the pointer $P$ and flips one of the high-order bits such that $P$ becomes invalid.
If the invalid pointer is used by an instruction that causes the pointer to be translated, such as load or instruction fetch, the \glsdesc{mmu} issues a memory translation fault.

\ifnotabridged
\Gls{pa} also supports the generic \pacgplain instruction, which outputs a 32-bit PAC based on a 64-bit input value and a 64-bit modifier. There is no corresponding verification instruction.
To verify the  \pacgplain \gls{pac}, instrumented code must explicitly compare it to the expected value.
\fi

\subsubsection{\Glstext{pa}-based return address protection}
\label{sec:pa-return-address-protection}
\Gls{pa}-based return address protection is implemented as part of the \branchProtFlag{} feature of GCC and LLVM/Clang.
\footnote{\url{https://gcc.gnu.org/gcc-9/changes.html} and \\\url{https://reviews.llvm.org/D51429}}
An authenticated return address is computed with \instr{paciasp} (\dOne in Listing~\ref{lst:qualcomm}) and verified with \instr{retaa} (\dFour).
These instructions use the instruction key $A$ and the value of \gls{sp} as the modifier.
The \gls{pa}-keys are protected by hardware; consequently an adversary has to resort to guessing the correct \gls{pac} for a modified return address.

\inputAsmListing{qualcomm}{
  The \branchProtFlag{} feature in GCC and LLVM/Clang uses \gls{pa} to sign (\dOne) and verify (\dFour) the return address in \gls{lr}.
  \Gls{pa} does not access memory directly, the \gls{lr} value is stored (\dTwo) and loaded (\dThree) conventionally.
}

The \branchProtFlag{} feature and other prior \gls{pa}-based solutions are
vulnerable to \emph{reuse attacks} where an adversary replaces a valid authenticated return address with another authenticated return address previously read from the process' memory.
For a reused \gls{pac} to pass verification, both the original and replacement \gls{pac}
must have been computed using the same \gls{pa} key and modifier.
This applies to any \gls{pa} scheme, not only authenticated return addresses.
\ifnotabridged{}For instance, if a constant modifier is used then all pointers based on the same key are interchangeable.\fi{}
Using the \gls{sp} value as a modifier reduces the set of interchangeable pointers, but still allows reuse attacks when \gls{sp} values coincide.
Reuse attacks can be mitigated, but not completely prevented, by further narrowing the scope of modifier values~\cite{Liljestrand19}.

\section{Adversary model and requirements}
\label{sec:threat_model}\label{sec:requirements}

In this work, we consider a powerful adversary, \emph{\theAttacker}, with arbitrary control of process memory but restricted by a \rowPolicy{} policy that prevents modification of code pages.
This adversary model is consistent with prior work on run-time attacks~\cite{Szekeres2013}.
We limit \theAttacker to user space; thus \theAttacker cannot read or modify kernel-managed registers such as the \gls{pa} keys.

\noindent{}We make the following assumptions about the system:
\begin{enumerate}[label=\textbf{{A\arabic*}},ref=\textbf{{A\arabic*}}]
\itemsep0em

  \item\label{ass:dep}\textit{A \rowPolicy{} policy}
  protects code memory pages from modification by non-privileged processes.
  All major processor architectures, including ARMv8-A, support \rowPolicy.

  \item\label{ass:cfi}\textit{Coarse-grained forward-edge \gls{cfi}}
  that restricts forward control-flow transfers to a set of valid targets.
  Specifically, we assume that indirect function-calls always target the beginning of a function and that indirect jumps to arbitrary addresses is infeasible.
  This property is satisfied by several pre-existing software-only \gls{cfi} solutions with reasonable overhead~\cite{Abadi09,Davi12,Kuznetsov14,Mashtizadeh15}, as well as with negligible overhead by using hardware-assisted mechanisms like ARM \gls{pa}~\cite{Liljestrand19}, branch target indicators~\cite{ARMv8A}, or TrustZone-M~\cite{Nyman17,ARMv8M}.
  In particular, a minimal \gls{pa} scheme using a constant (e.g., \texttt{0x0}) modifier fulfills this assumption.

\end{enumerate}

This adversary model allows \theAttacker{} to modify any pointer in data memory pages.
In particular, \theAttacker{} can modify function return addresses while they reside on the program call stack.
\ref{ass:cfi}~and~\ref{ass:dep} prevent \theAttacker{} from tampering with \SHORTNAME{} instrumentation (Section~\ref{sec:evaluation-security-runtime}).
Our goal is to thwart \theAttacker{} who modifies function return addresses in order to hijack the program control flow.
We define the following requirements:

\begin{enumerate}[label=\textbf{{R\arabic*}},ref=\textbf{{R\arabic*}}]
\itemsep0em

  \item\label{req:auth}\textit{Return address integrity}: Detect if a function return address has been modified while in memory.

  \item\label{req:reuse}\textit{Memory disclosure tolerance}: Remain effective even when \theAttacker{} can read the entire process address space.

  \item\label{req:comp}\textit{Compatibility}: Be applicable to typical (standard-compliant) C code without source code modifications.

  \item\label{req:perf}\textit{Performance}: Impose only minimal run-time performance and memory overhead, while meeting~\ref{req:auth}--\ref{req:comp}.

\end{enumerate}
As in prior work on \gls{cfi}, we do not consider non-control data attacks~\cite{Chen05}, such as \gls{dop}~\cite{Hu16}.

\begin{figure*}[tp]
\centering
\ifbw{}\includegraphics[width=0.9\textwidth]{acs-concept-bw}
\else\includegraphics[width=0.9\textwidth]{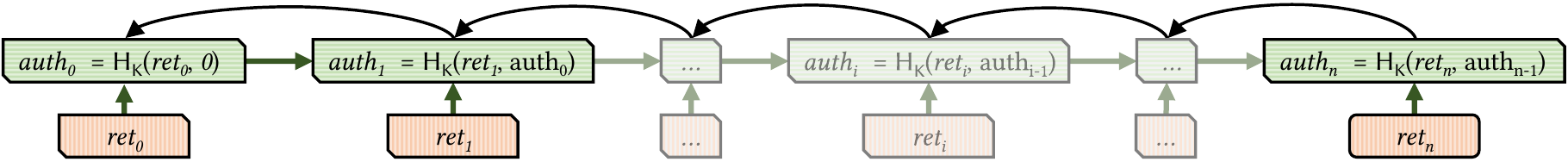}\fi
\caption{
\SHORTNAME{} is an chained \gls{mac} of tokens $\auth{i}, i\in[0,n-1]$ that are cryptographically bound to the corresponding return addresses, $\ret{i}, i\in[0,n]$, and
the last \auth{n}.
}
\label{fig:acs-concept}
\end{figure*}

\section{Design: \LONGNAME}
\label{sec:design}\label{sec:design-pacga}

\begin{figure}[tp]
\centering
\ifbw{}\includegraphics[width=1\columnwidth]{call-stack-split-bw}
\else\includegraphics[width=1\columnwidth]{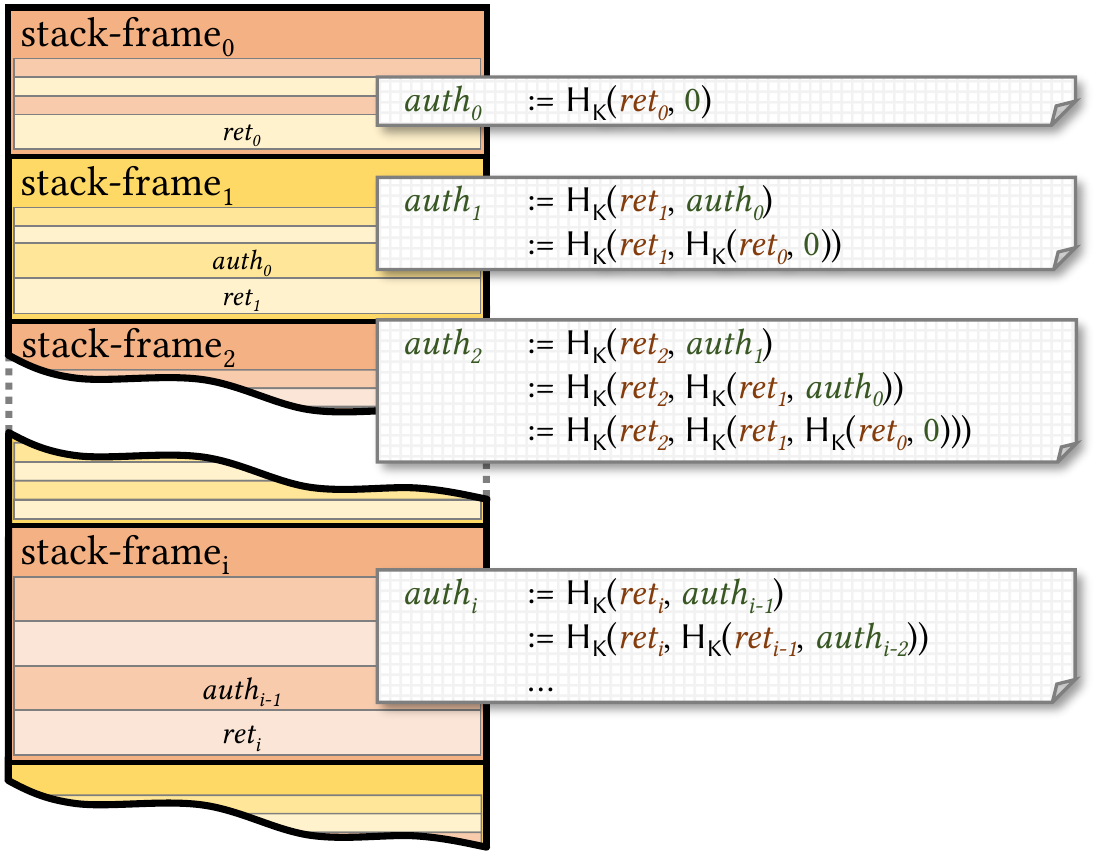}\fi
\caption{
\SHORTNAME{} stores return addresses and intermediate authentication tokens, $auth_i, i \in [0, n-1]$, on the stack.
  Only the last token ($auth_n$) needs to be securely stored.
}
\label{fig:call-stack}
\end{figure}

In this section we present our general design for an \LONGNAME (\SHORTNAME)\ifnotabridged{}, not tied to a particular hardware-assisted mechanism\fi{}.
In Section~\ref{sec:implementation}, we present our implementation that efficiently realizes \SHORTNAME using ARM \gls{pa}.
Our key idea is to provide a modifier for the return address by cryptographically binding it to \emph{all previous return addresses in the call stack}.
This makes the modifier statistically unique to a particular control-flow path, thus preventing reuse-type attacks and allowing \emph{precise verification of return addresses}.
The return addresses $ret_i,i\in[0,n]$ (where $n$ is the depth of the call stack in terms of active function records) must be stored on the stack, where \theAttacker can modify them by exploiting memory vulnerabilities.
\SHORTNAME{} protects these values by computing a series of \emph{chained}
authentication tokens $auth_i,i\in[0,n]$ that cryptographically bind the last
$auth_n$ to all return addresses
$ret_i,i\in[0,n-1]$ stored on the stack (Figure~\ref{fig:acs-concept}).
Only the \gls{mac} key and the last authentication token $auth_n$ must be stored securely to ensure that previous $auth$ tokens and return addresses can be correctly verified when unwinding the call stack (\ref{req:auth}).
We use a tweakable \gls{mac} function $H_{K}$ to generate a $b$-bit authentication token $auth_i$:
\[ auth_i =
  \begin{cases}
    \pachash{ret_{i}}{auth_{i-1}}  & \quad \text{if } i > 0\\
    \pachash{ret_{i}}{0}           & \quad \text{if } i = 0\\
  \end{cases}
\]
$auth_n$ is maintained in a register unmodifiable by \theAttacker.
Figure~\ref{fig:call-stack} shows how authentication tokens and return addresses are stored on the call stack.
On function calls, $auth_i$ is retained across the call to the callee, which calculates $auth_{i+1}$ and stores both $auth_i$ and the corresponding return address $ret_{i+1}$ on its stack frame.
On return, $auth'_{i-1}$ and $ret'_i$ values are loaded from the stack and are verified by comparing $\pachash{auth'_{i-1}}{ret'_i}$ to $auth_i$.
If the results differ, then one or both of the loaded values have been corrupted (\ref{req:auth}).
Otherwise, they are valid---i.e., $\authAdv{i-1} = \auth{i-1}$ and $\retAdv{i} = \ret{i}$---in which case \auth{i} is replaced with the verified \auth{i-1} in the secure register before the function returns to \ret{i}.

For compactness, we can combine $auth_i$ and $ret_i$, into an \emph{authenticated
  return address}, $aret_i$:
\begin{align*}
  aret_i &= auth_i \parallel ret_i \mathrm{, where} \\
  auth_i &=
           \begin{cases}
             \pachash{ret_{i}}{aret_{i-1}}  & \quad \text{if } i > 0\\
             \pachash{ret_{i}}{0} & \quad \text{if } i = 0\\
           \end{cases}
\end{align*}
We call $auth_{i}$ and the corresponding $aret_i$ \emph{valid} if
$auth_i = \pachash{ret_{i}}{aret_{i-1}}$ for some given $aret_{i-1}$.

\subsection{Securing the authentication token}
\label{sec:design-securing-token}

The current authenticated return address $aret_n$, is secured by keeping it exclusively in a CPU register which we call the \emph{\gls{cr}}.
Note that reserving exclusive use of a register is also a requirement for current shadow stack implementation for the 64-bit ARM architecture~\cite{ShadowCallStack} and has been proposed for shadow stacks on the x86 architecture~\cite{Burow19}.

\SHORTNAME{} protects the integrity of backward-edge control-flow transfers. Combined with coarse-grained forward-edge \gls{cfi} (Assumption~\ref{ass:cfi}), it ensures that:
\begin{enumerate*}[label=\arabic*)]
\item\label{design-req:authRet} immediately after function return, the $aret_n$ in \gls{cr} is valid,
\item\label{design-req:authPro} at function entry the $aret_{n-1}$ stored in \gls{cr} is valid, and
\item\label{design-req:lr} \gls{cr} is always used as or set to a valid $aret$.
\end{enumerate*}
This ensures that token updates are done securely, and that the \SHORTNAME{}
instrumentation cannot be bypassed or used to generate arbitrary
authenticated return addresses.

\ifnotabridged\subsection{Mitigating hash-collisions: authentication token masking}
\else\subsection{Mitigating hash-collisions}\fi
\label{sec:design-collisions}

Though $aret_n$ is protected by hardware, the size $b$ of the authentication token $auth$ can be limited by the implementation.
Using a \gls{pac} as the token would typically limit it to 16 bits.
This is significant, as collisions can be found after \theAttacker has seen, on
average, approximately $1.253 \cdot 2^{b/2}$ tokens~\cite[Section~1.4.2]{Smart16} (e.g., 321 tokens for $b=16$).
Despite this, we can still prevent \theAttacker{} from \emph{recognizing}
collisions (\ref{req:reuse}), thus forcing \theAttacker{} to guess---with a success probability $2^{-b}$---which authenticated return
addresses yield a collision.
The $auth$ of any $aret$ stored on the stack is masked using a pseudo-random
value derived from the previous $aret$ value:
\begin{align*}
  auth_{i} &= \pachash{ret_i}{aret_{i-1}} \oplus \pacmask{aret_{i-1}} .
\end{align*}
The mask is exclusive-OR-ed with
$\pachash{ret_i}{aret_{i-1}}$ after it is generated and before it is authenticated, thereby preventing \theAttacker from identifying opportunities for
pointer reuse.
We discuss the security of masking in~\Cref{sec:evaluation-pac-collision}.

\ifnotabridged\subsection{Mitigation brute-force guessing: re-seeding authentication token chain}
\else\subsection{Mitigating brute-force guessing}\fi
\label{sec:brute-force-mitigation}

A brute force attack where \theAttacker guesses an $auth$ token succeeds with probability $p$ for a $b$-bit $auth$ after $\frac{\log(1-p)}{\log(1-2^{-b})}$ guesses, provided that a failed guess terminates the program and subsequent program runs use a new key to generate $auth$ tokens.
This assumption is similar to prior \gls{pa}-based solutions~\cite{Liljestrand19} and is consistent with current \gls{pa} behavior in Linux 5.0.
However, if pre-forked or multithreaded programs share the key, \theAttacker can target a vulnerability in a sibling.
Unless a failed authentication terminates the entire process tree, \theAttacker can then attempt a new guess against another sibling process without resetting the key.
In this scenario, $2^{b-1}$ guesses on average are enough to obtain a modifier with respect to which some combination of pointer and authentication token is valid.
Since this modifier becomes the next authenticated return address, the process can be repeated to use the injected address.
Because the two guesses can be done separately using a divide-and-conquer strategy, this requires on average $2^b$ guesses to allow \theAttacker to jump to an arbitrary address, rather than $2^{2b}$ that are needed when the guesses are independent.

\ifnotabridged
Multi-threaded applications are also affected since address translation errors due to PAC authentication failures are delivered in Linux via the \texttt{SIGSEGV} signal which is always directed against the offending thread\footnote{\url{http://man7.org/linux/man-pages/man7/signal.7.html}}, and the thread cannot change the signal's disposition such that it would not be delivered. 
\fi

Liljestrand et al.~\cite{Liljestrand19} recommend hardening pre-forking and multi-threaded applications against guessing attacks by having the application restart all of its processes if the number of PAC failures in child processes exceeds a pre-defined threshold.
We recommend an alternative mitigation specific to \SHORTNAME: \emph{"re-seeding"} the $auth$ calculation after a fork or thread creation.
For example, calculating $auth_{0} = \pachash{ret_0}{init}$ where $init$ corresponds to the process or thread ID.
This solution is straightforward to apply to threads, as a return from the function starting the thread causes the thread to exit.
Crucially, re-seeding prevents a divide-and-conquer guessing strategy and requires on average $2^{2b}$ guesses.
Therefore, the \SHORTNAME for the thread stacks can be made disjoint from the main \SHORTNAME chain.
However, forked processes may use $auth$ tokens in stack frames inherited from the parent process.
If a child process never returns to inherited stack frames, re-seeding any new $auth$ tokens beyond the point of the fork is sufficient.
However, if the child process returns to inherited stack frames, the \SHORTNAME must be re-seeded starting from $auth_0$ by rewriting any $auth$ tokens in pre-existing stack frames; similar to some stack canary re-randomization schemes~\cite{Petsios15,Hawkins16}.

\subsection{Irregular stack unwinding}
\label{sec:design-unwind}

The C standard includes the \setjmp~/ \longjmp programming interface, which can be used to add exception-like functionality to C\ifnotabridged~(Listing~\ref{lst:longjmp})\fi.
The \longjmp C function executes a non-local jump to a prior calling environment stored using the \setjmp function.
At \setjmp, callee-saved registers (whose values are guaranteed to
persist through function invocations), as well as the stack pointer
\gls{sp}, and the return address are stored in the given \jmpbuf buffer\ifnotabridged~(\dCOne in Listing~\ref{lst:longjmp})\fi.
\ifnotabridged
\setjmp returns \texttt{0} to indicate that execution is continuing directly after the call.
Upon executing \longjmp, the environment is restored from \jmpbuf (\dCThree); program execution continues at the \setjmp return site with a non-zero value (\dCTwo).\fi
Calling \longjmp{} using an expired buffer, i.e., after the corresponding
\setjmp{} caller has returned\ifnotabridged~(\dCFour)\fi, results in
undefined behavior (the implications of this are discussed in Section~\ref{sec:discussion-unwind}).
Because \jmpbuf also stores the last authenticated token, \SHORTNAME needs a
mechanism to ensure its integrity when using \setjmp and \longjmp.

\ifnotabridged
\begin{lstlisting}[float,floatplacement=tp,style=customc,label={lst:longjmp},linerange={2-100},
caption={
\setjmp~/ \longjmp allows the programmer to transfer execution to another location, potentially in another function.
The location, and the state of the environment after the transfer, is determined by an in-memory buffer containing the calling environment of a previous \setjmp call.
Calling \longjmp after the calling environment is destroyed results in undefined behavior.
}]
#include <setjmp.h>

jmp_buf ebuf;

void try_catch() {
  int err;

  if (!(err = setjmp(ebuf))) { // iff ebuf set (*@ \hfill\dCOne@*)
    checked_func();            // after setjmp
  } else { 
    handle_error(err);         // after longjmp (*@ \hfill\dCTwo@*)
  }
}

void checked_func() {
  // ...
  longjmp(ebuf, E_NUM); // throw exception (*@ \hfill\dCThree@*)
  // ...
}

int main() {
  try_catch();
  longjmp(ebuf, E_NUM); // undefined behavior! (*@ \hfill\dCFour@*)
}
\end{lstlisting}
\fi

While in memory, the integrity of \jmpbuf cannot be guaranteed.
Nonetheless, the stored \auth{i} is bound to the corresponding \auth{i-1} on the \setjmp caller's stack.
This ensures that \longjmp always restores a valid \SHORTNAME state.
To limit the set of values \theAttacker{} can inject into \jmpbuf, we replace the \setjmp return address \ret{b} in \jmpbuf with \aret{b}, defined as:
\[
aret_{b} = \left(\pachash{ret_{b}}{auth_{i}} \parallel ret_{b} \right) \oplus \pachash{\mReg{SP}_b}{auth_{i}},
\]
where $\mReg{SP}_b$ is the \gls{sp} value stored in \jmpbuf.
When executing \longjmp{}, $aret_b$ is recalculated based on the buffer values to verify that the stored $auth_{i}$ was stored by a \setjmp.
\theAttacker{} cannot generate the $aret_{b}$ value for an arbitrary $auth_i$, nor replace $aret_{b}$ with a previously observed $auth_i$.
But, since \longjmp explicitly allows jumping to prior states, \SHORTNAME cannot ensure that the target is the \emph{intended one}, i.e., \theAttacker could substitute the correct \jmpbuf with another.
Shadow stacks share a similar limitation~\cite{Dang15}, and cannot guarantee that the intended state has been reached, only that the return address (and stack pointer) in that state is intact.

\section{Implementation: \IMPLNAME}
\label{sec:implementation}

\glsreset{cr}
\glsreset{va}

We present \IMPLNAME, an \SHORTNAME realization using \ARMPAFULL.
\IMPLNAME is based on LLVM 9.0 and integrated into the 64-bit ARM backend.
\IMPLNAME modifies the \texttt{AArch64FrameLowering} such that the function stores and loads \aret{n-1} during \texttt{FrameSetup} and \texttt{FrameDestroy}, respectively.
We also modify the \texttt{AArch64RegisterInfo} to ensure that the register holding \aret{n}, \CR, is reserved for \IMPLNAME use.
Our current implementation uses a \gls{ir} pass to mark all functions for instrumentation, whereas the backend then performs instrumentation based on the function attribute.
\ifnotabridged
The instrumentation is controlled \texttt{-pacstack=}, which is passed to the compiler backend using the \texttt{-mllvm} flag in Clang.
We plan to replace this approach with direct support in the Clang frontend.
\fi
\ifanonymous{}We plan to open source our compiler modifications.\footnote{\IMPLNAME source code is available in the supplementary material at \\\material}
\fi

\ifnotabridged{}
\subsection{Authenticated return addresses with \gls{pa}}

\todo{HL: This needs to be updated!}
\textbf{Variant 1: generating $auth$ with \pacgplain.}
\label{sec:implementation-pacga}
\label{sec:implementation-autia}
In this variant, we use \pacgplain to generate $auth$ tokens:
\[ \mReg{Xd} \leftarrow \auth{i} =
  \begin{cases}
  \pacg{\mReg{Xd}, \mReg{LR}=ret_{i}}{\mReg{CR}=\auth{i-1}} & \quad \text{if } n > 0\\
  \pacg{\mReg{Xd}, \mReg{LR}=ret_{i}}{\mReg{CR}=init}        & \quad \text{if } n = 0\\
  \end{cases}
\]
To generate and verify authentication tokens, \IMPLNAME instruments function prologues and epilogues (Listing~\ref{lst:acs-pacga}).
In the function prologue, $auth_{i-1}$ and $ret_i$ (in \gls{cr} and \gls{lr}, respectively) are stored on the function stack frame and then used to generate a new $auth_i$ with \pacgplain (\dOne).
The \auth{i-1} and \ret{i} values are then stored on the function stack frame.
Before function return, \IMPLNAME verifies the \authAdv{i-1} and \retAdv{i} read from the stack by calculating the corresponding \authAdv{i} (\dThree) and comparing it to \auth{i}, stored in \gls{lr} (\dFour).
For $auth_0$ any value currently in \gls{cr} is used and stored for later validation.
This allows \IMPLNAME to operate without explicit initialization by the C Library (\texttt{libc}) startup code.
To enable re-seeding the $auth$ token chain (Section~\ref{sec:brute-force-mitigation}), the process and thread initialization, and \texttt{fork()} wrapper in \texttt{libc} should be modified to set the initial value of \texttt{CR} accordingly.

\inputAsmListing{acs-pacga}{
Variant~1 of \IMPLNAME generates and verifies $auth$ tokens using \instr{pacga} (\dOne and \dThree).
Both \auth{i-1} and \ret{i} are stored on the stack, and are hence validated against \auth{i} on function return (\dTwo).
Where possible, the \emph{store pair} (\instr{stp}) / \emph{load pair} instructions (\instr{ldp}) are used to minimize the latency for successive loads / stores.
}

Variant~1 can efficiently compute 32-bit authentication tokens values using \pacgplain.
However, it has two drawbacks:
First, an additional stack store~/ load is added for the 4-byte token;
to preserve the callee-saved behavior of \gls{cr}, the full 8-byte register content must be stored on the stack.
Second, the output of \pacgplain must be explicitly checked using a comparison and a conditional branch instruction.
For this reason, our current implementation only supports variant~2 below.

However, in Section~\ref{sec:discussion-generalizing-acs} we discuss using \pacgplain to bind other stack-based write-once data to a specific \SHORTNAME state.

\textbf{Variant 2: generating $aret$ with \autiplain.}

Variant-1 follows our generic design and maintains a separate authentication token.
In variant-2, we instead
\else{}
\fi{}

The current authenticated return address is securely stored in \gls{cr}.
Because the unprotected return address $ret_i$ is never stored on the stack, \theAttacker{} is limited to manipulating the earlier authenticated return addresses on stack, i.e., $aret_{i}, i\in[0,n-1]$.
An authenticated return address must therefore pass two authentications
before use: first when being restored from the stack, and second, when being used
as the target of a function return.
We discuss the security implications in Section~\ref{sec:evaluation-security}.

\IMPLNAME uses the \paciplain and \autiplain instructions to efficiently calculate and verify authenticated return addresses (Listing~\ref{lst:acs-fast}, \acsFastProPac and \acsFastEpiAut).
\ifnotabridged
These instructions differ from \pacgplain in that the output
\else
The result of \paciplain
\fi
is $aret_{i}$ which is stored in the \glsdesc{lr} (\glstext{lr}, \acsFastProPac) and moved to \gls{cr} (\acsFastProCR):
\begin{align*}
\mReg{LR} \leftarrow \aret{i} =
\begin{cases}
    \paci{\mReg{LR}=\ret{i}}{\mReg{CR}=\aret{i-1}}  & \quad \text{if } i > 0\\
    \paci{\mReg{LR}=\ret{i}}{\mReg{CR}=init}         & \quad \text{if } i = 0\\
  \end{cases}
\end{align*}
The corresponding verification (\acsFastEpiLdrAut and \acsFastEpiAut) are defined as:
\begin{align*}
\mReg{LR} \leftarrow \auti{\mReg{LR}=\aret{i}}{\mReg{CR}}  =
\begin{cases}
    \ret{i}    & \quad \text{if } \pachash{ret_i}{\mReg{CR}} = auth_i  \\
    \retBad{i} & \quad otherwise, \\
  \end{cases}
\end{align*}
where \autiplain will automatically handle verification errors by setting
\gls{lr} to an unusable address \retBad{i}.
No additional checking is needed; executing a return to \retBad{i} causes a address translation fault (Section~\ref{sec:background-pa}).
To maintain compatibility (\ref{req:comp}), \IMPLNAME does not modify the frame record (\acsFastProStrFrameRecord) and instead stores $aret_{i-1}$ in a separate stack slot (\acsFastProStrAut).
This allows, for instance, debuggers to backtrace the call-stack without knowledge of \IMPLNAME.
\IMPLNAME never loads $ret_{i}$ from the frame record; it always uses $aret_{i}$ which is securely stored in \mReg{CR}.

\inputAsmListing{acs-fast}{
  At function entry, \IMPLNAME stores \aret{i-1} on the stack (\acsFastProStrAut) and generates a new \aret{i} (\acsFastProPac) which is retained in \mReg{CR} (\acsFastProCR).
Before return, \aret{i-1} is loaded from the stack (\acsFastEpiLdrAut) and verified against \aret{i} (\acsFastEpiAut).
Verification failure sets \mReg{LR} to an invalid address \retBad{i} and causes a fault on return.
}

\subsection{Securing the authentication token}
\label{sec:implementation-securing-token}
\label{sec:function-call-instrumentation}

\IMPLNAME uses the ARM general purpose register \crDef as \gls{cr} for storing the last authentication token.
\crDef is a callee-saved register, and so, any function that uses it must also restore the old value before return.
By using \crDef as \gls{cr}, \IMPLNAME libraries or code can be transparently mixed with uninstrumented code (\ref{req:comp}).
We discuss the security implications of mixing instrumented and uninstrumented code in Section~\ref{sec:discussion-legacy}.

\subsection{Mitigating hash collisions: \gls{pac} masking}
\label{sec:auth-token-masking}

To prevent \theAttacker from identifying \gls{pac} collisions that can be reused
to violate the integrity of the call stack, \IMPLNAME masks all authentication tokens values before storing them on the stack.
A pseudo-random value is obtained by generating a \gls{pac} for address \texttt{0x0}, $\paci{0}{aret_{i-1}}$ (Listing~\ref{lst:acs-mask} \acsMaskProPacMask, \acsMaskEpiPacMask).
By using \paciplain we efficiently obtain a pseudo-random value that can be
directly applied to the authentication token part of $aret$ using only an
exclusive-or instruction (\instr{eor} \acsMaskProEorMask, \acsMaskEpiEorMask).

Because this construction uses the same key to generate both authentication tokens and masks, \theAttacker must not obtain an \aret{i} for a $\ret{i}=\mathtt{0x0}$ and any existing \aret{i-1}.
\IMPLNAME will never generate such $aret$ values, as the return address never points to memory address zero.
To prevent leaking the mask directly, it is cleared after use (\acsMaskProMaskClear, \acsMaskEpiMaskClear).
Consequently no \pachash{0}{x} value is visible to \theAttacker
nor is it possible to pre-compute without the confidential \gls{pa} key.

This approach to masking requires two additional \gls{pac} calculations for each function activation.
\IMPLNAME supports instrumentation with or without masking\ifnotabridged via the \texttt{-pacstack=full} and \texttt{-pacstack=nomask} flags respectively\fi.
We discuss the security of \gls{pac} masking in Section~\ref{sec:evaluation-pac-collision}.

\inputAsmListing{acs-mask}{
\IMPLNAME masks authentication tokens to hide collisions.
The mask is created with $\paci{0}{\aret{i-1}}$ (\acsMaskProPacMask), and exclusive-OR-ed with the unmasked \auth{i} (\acsMaskProEorMask).
On return, the masked \auth{i} is loaded from the stack (\acsMaskEpiLdrAut).
The mask is then recreated (\acsMaskEpiPacMask) and removed from \auth{i} (\acsMaskEpiEorMask) before verification.
\mReg{X15} is a scratch register and can be safely used as its value is not retained between function calls.
}

\subsection{Irregular stack unwinding}
\label{sec:implementation-stack-unwind}

\IMPLNAME binds \jmpbuf buffers to the $aret_i$ at the time of \setjmp call
by replacing the \setjmp return address \ret{b} with its authenticated counterpart $aret_b$ before \setjmp stores it to the \jmpbuf (Section~\ref{sec:design-unwind}).
The \libc implementation is not modified; instead \setjmp~/ \longjmp calls are replaced with the wrapper functions in Listings~\ref{lst:setjmp_wrapper}~and~\ref{lst:longjmp_wrapper}.

The \setjmpWrapper (Listing~\ref{lst:setjmp_wrapper}) replaces the return address in \mReg{LR} with \aret{b} and then executes \setjmp, which stores it in the buffer.
The \longjmpWrapper (Listing~\ref{lst:longjmp_wrapper}) retrieves $aret_b$, $aret_i$, and the \gls{sp} values from \jmpbuf, verifies their values and writes $ret_b$ into \jmpbuf before executing \longjmp.

\inputAsmListing{setjmp_wrapper}{
  \IMPLNAME redirects \setjmp calls to our \func{setjmp\_wrapper}\!\protect\footref{fn:wrapper-code} which binds the return address \aret{b} to \aret{i} and the \gls{sp} value before it is stored in \jmpbuf.
}

\inputAsmListing{longjmp_wrapper}{
  ~~Before \longjmp,~the~\IMPLNAME \func{longjmp\_wrapper}\!\protect\footref{fn:wrapper-code} verifies the binding of the \aretAdv{b}, \retAdv{b} and $sp'_b$ values stored in \jmpbuf.
\theAttacker cannot generate \aretAdv{b} for arbitrary values and therefore cannot inject them in \jmpbuf.
\texttt{\#r}, \texttt{\#a} and \texttt{\#s} are the offsets to \ret{b}, \gls{cr}, and \ret{i} within \jmpbuf.
}

\footnotetext{\label{fn:wrapper-code}Listings~\ref{lst:setjmp_wrapper}~and~\ref{lst:longjmp_wrapper} are illustrative, complete wrapper code is available at \url{https://github.com/pacstack/pacstack-wrappers}}

\subsection{Multi-threading}
\label{sec:implementation-multithreading}

The values of ARMv8-A general purpose registers are stored in memory when entering \gls{el}1 (i.e. kernel-mode) from \gls{el}0
(i.e. user-mode), for example during context switches and system calls.
This must not allow \theAttacker to modify the $aret$ values or read the mask, which
are both exclusively in either \gls{cr} or \gls{lr} during execution (Listings~\ref{lst:acs-fast} and~\ref{lst:acs-mask}), but must be stored in memory during the context switch.
On ARMv8-A, system calls are implemented using the supervisor call instruction (\instr{svc}) that switches the CPU to \gls{el}1 and triggers a configured handler.
On 64-bit ARM, Linux \version{5.0} uses the \texttt{kernel\_entry}\footnote{\url{https://git.kernel.org/pub/scm/linux/kernel/git/torvalds/linux.git/tree/arch/arm64/kernel/entry.S?h=v5.0}} macro to store all register values on the \gls{el}1 stack, where they cannot be accessed by user-space processes.
During context switches, callee-saved registers (including \gls{cr}) and \gls{lr} are stored in \texttt{struct cpu\_context}\footnote{\url{https://git.kernel.org/pub/scm/linux/kernel/git/torvalds/linux.git/tree/arch/arm64/include/asm/processor.h?h=v5.0}} which belongs to the in-kernel task structure and cannot be accessed by user space.
The \gls{cr} and \gls{lr} values of a non-executing task are thus securely stored within the kernel, beyond the reach of other processes or other threads within the same process.
Thus, no kernel modifications are needed to securely apply \IMPLNAME to multi-threaded applications.

\section{Security evaluation}
\label{sec:evaluation-security}

We address three questions in this section:\\
\begin{inparaenum}[1)]
\item Is \gls{pac} reuse a realistic concern in prior \gls{pa}-based schemes?
\item Is the \SHORTNAME scheme cryptographically secure?\\
\item Do \SHORTNAME's guarantees hold when instantiated as \IMPLNAME?
\end{inparaenum}

\subsection{Reuse attacks on \gls{pa}}
\label{sec:evaluation-security-reuse-in-practice}

Reuse attack on PA-based schemes are possible when the modifier is calculated with known or predictably repeating values.
Using the \gls{sp} can mitigate reuse attacks (\Cref{sec:pa-return-address-protection}).
However, \branchProtFlag{} generates the \gls{pac} immediately on function entry, before modifying the \gls{sp} value to allocate stack space.
All functions called from within a code segment use the same modifier unless there are dynamic stack allocations.
Moreover, because the stack is typically aligned to 8 bytes, the \gls{sp} value will often repeat.
For example, a less than $1s$ test execution of a SPEC CPU 2017 benchmark (\texttt{538.imagick\_r}) already shows multiple collisions, with $5349$ distinct (\gls{lr},\gls{sp}) pairs, but only $914$ unique \gls{sp} values.
\Cref{lst:attack_example} shows a minimal example where all called functions will end up using the same modifier and thus have interchangeable signed return addresses.

\inputCListing{attack_example}{
The \branchProtFlag{} implementation (\Cref{sec:pa-return-address-protection}) computes the \gls{pac} for return addresses using the \gls{sp} value at function entry.
Both invocations in \texttt{func} (\Cref{line:attack_example:call_a,line:attack_example:call_b}) will thus use the same \gls{sp} value as modifier.
\theAttacker can reuse the signed address from \Cref{line:attack_example:call_a} to make the function invocation at \Cref{line:attack_example:call_b} return to \Cref{line:attack_example:bad_return}.
}

\subsection{\SHORTNAME security}
\label{sec:evaluation-security-reuse}

\begin{figure*}
  \centering
  \begin{subfigure}[t]{0.4\linewidth}
    \raisebox{1.062cm}{\includegraphics[height=3.5cm]{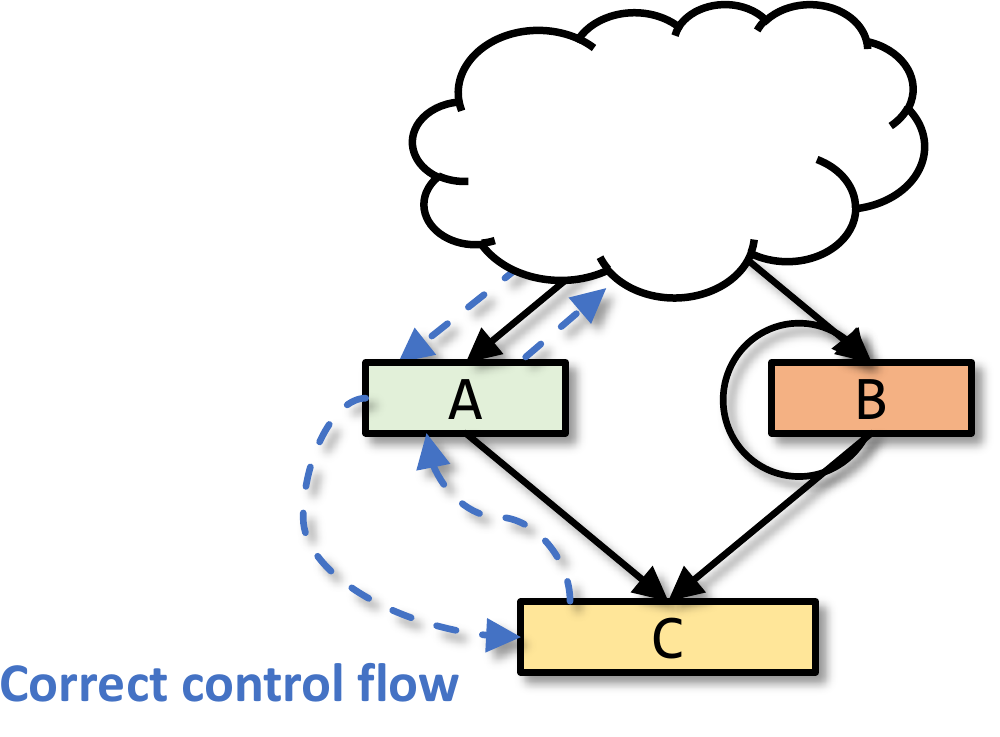}}
    \caption{Normal control flow.}
    \label{fig:call-graph-normal}
  \end{subfigure}
  \begin{subfigure}[t]{0.4\linewidth}
    \includegraphics[height=4.562cm]{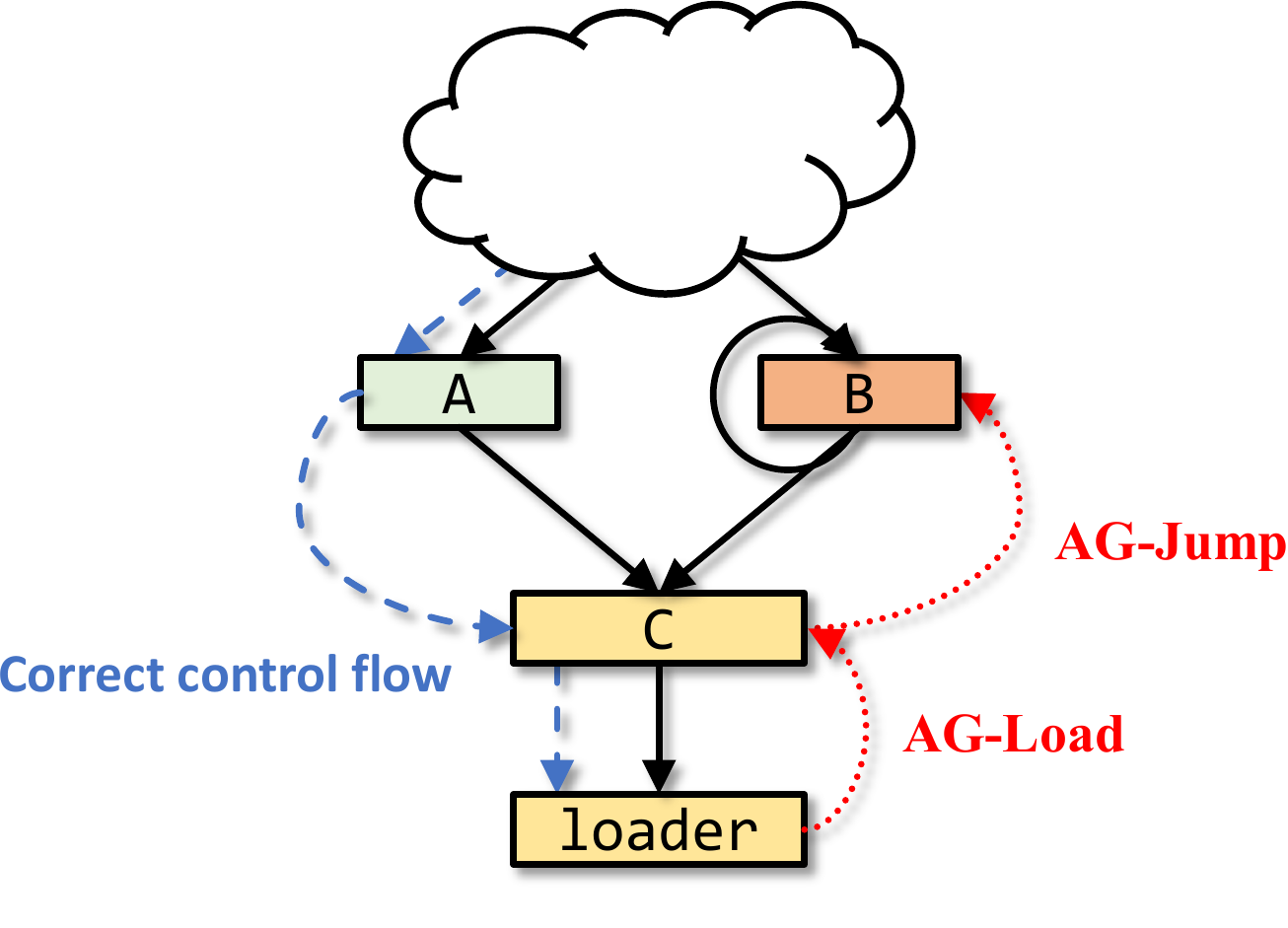}
    \caption{\theAttacker's desired control flow.}
    \label{fig:call-graph-collision}
  \end{subfigure}
  \caption{Anatomy of a backward-edge control-flow attack against \SHORTNAME. In
    order to force function $C$ to return to $B$ instead of its caller $A$,
    \theAttacker substitutes their authenticated return address \aret{B} when
    some function---the `loader'---returns to \ret{C} in function $C$ (goal
    AG-Load). If \aret{B} is valid with respect to some known modifier, then at
    the end of function $C$ the program will return to the corresponding \ret{B}
    (goal AG-Jump).}
  \label{fig:call-graph}
\end{figure*}

A generic representation of an attack against \SHORTNAME is shown in
Figure~\ref{fig:call-graph}.
Under normal operation, function $C$ returns to $A$ if called from $A$ (Figure~\ref{fig:call-graph-normal}); i.e., when called from $A$, the return address of $C$ is an address \ret{A} in $A$.
The goal of \theAttacker (Figure~\ref{fig:call-graph-collision}) is to cause $C$ to return to
some other address $ret_B$.

Since the authenticated return address $aret_A$ containing $ret_A$ is protected from \theAttacker,
in order to perform a backward-edge control-flow attack, \theAttacker must achieve two
goals successfully: \\

\textbf{AG-Jump:} Obtain an authenticated return address $aret_B$, valid with
respect to some known modifier, which will validate successfully when
$C$ returns. \\

\textbf{AG-Load:} Violate the integrity of the call stack such that
  the \gls{lr} register is loaded with $aret_B$ from AG-Jump rather than the
  correct authenticated return address
  $aret_A$. \\

This requires two returns: one from a `loader' function to load \theAttacker's
$aret_B$ into \gls{lr}, and another from $C$ to the return address $ret_B$ contained in $aret_B$.

In the analyses below, we treat the $auth$ token \pachash{$P$}{m} as a random
oracle with respect to both the pointer $P$ and modifier $m$.  This means that
if \pachash{P}{m} has never been computed by a function call, \pachash{P}{m}
will match any value with probability $2^{-b}$, independently of any other value
\pachash{P'}{m'}.
In the analysis below we assume that programs that share the same PA keys between multiple processes or threads employ the mitigation strategy against brute-force attacks described in Section~\ref{sec:brute-force-mitigation}.
This assumption and the design of \SHORTNAME ensure that there
is no authentication oracle available: the only way to test whether an
$auth$ token is valid with respect to some address and modifier is to
attempt to return using the address and token, triggering a crash if the token is incorrect.
The difficulty of achieving these goals therefore depends on whether \theAttacker's
desired control-flow violation follows the call graph of the program and whether
$auth$ tokens are masked.  Violating
control-flow integrity while still traversing the call graph is easier because this
allows \theAttacker to harvest $auth$ tokens and search for collisions; violations
that do not follow the call graph are more difficult because they require that
\theAttacker make one or more guesses, risking a crash.

\subsubsection{Violations that follow the call graph}\label{sec:evaluation-pac-collision}

As \theAttacker can harvest authenticated return pointers when they are written to the stack,
the short $auth$ tokens mean that in the absence of masking an attacker
can violate the integrity of the call stack by finding collisions in $\pachash{\cdot}{\cdot}$.

In order to achieve goal AG-Load, \theAttacker must find two authenticated return
addresses \aret{A} and \aret{B}, such that \begin{inparaenum}[i)]
\item they are both returned to by a function $C$,
\item that $C$ contains a call-site to the loader function with a corresponding return address \ret{C}, and
\item such that
\end{inparaenum}
\begin{align}
   \pachash{ret_C}{aret_A} = \pachash{ret_C}{aret_B} = auth_\mathrm{collision}. \label{eqn:collision}
\end{align}

Note that the collisions must be for different values in the second argument
only, since that is the value in \theAttacker's control. Collisions that
require different values for $ret_C$ cannot be exploited because \ret{C}
is in \gls{cr} and cannot be modified by \theAttacker.

The $auth$ tokens contained in $aret_A$ and $aret_B$ depend on the path that
\theAttacker has taken through the call graph. \theAttacker can obtain as
many $auth$ tokens with $ret_C$ as a pointer as there are
distinct execution paths leading to $C$. The number of such
paths will explode combinatorially as the complexity of the program increases,
and cycles in the call graph---as occur in Figure~\ref{fig:call-graph}---make the number of paths essentially
infinite, limited only by available stack space.

Having found such a collision, \theAttacker then arranges for function $C$
to be called, traversing the call graph in such a way that it is set up to
return to $A$ using $aret_A$. Then, when the function $C$ calls into the loader function,
it will set \gls{lr} to $aret_{C}$. When the loader function
returns to $ret_{C}$, it will attempt to load $aret_A$ from the stack.
Instead, \theAttacker substitutes $aret_B$, which because of (\ref{eqn:collision})
will validate correctly when returning to $ret_\textsc{C}$. Since $aret_B$ is a
valid authenticated return address, $C$
will successfully return to $ret_B$, thereby violating the integrity of the call
stack.

More concretely, after collecting $q$ $auth$ tokens, according to the birthday
paradox~\cite[Section~1.4.2]{Smart16}, the probability
that \emph{some} pair collides
is:
\begin{align*}
  p_\textrm{collision}(q) &= 1-\frac{2^{b}!}{(2^{b}-q)! \cdot 2^{q \cdot b}}
\end{align*}
This quickly approaches $1$ as \theAttacker collects more tokens, on average
occurring after obtaining \[q = \sqrt{\frac{\pi 2^{b}}{2}}\] tokens. With a
$16$-bit \gls{pac}, \theAttacker will therefore obtain a collision after
harvesting $321$ pointers on average.

In order to successfully mount the above attack, \theAttacker must find two colliding $auth$ tokens and perform the substitution.
Without masking, \theAttacker can read the $auth$ token from the stack.
\theAttacker
can then keep collecting $auth$
tokens until they find two that collide; since these are both valid pointers,
\theAttacker will always succeed once this occurs, thus
\[\PP[\text{AG-Load} | \textrm{Collision}] = 1. \]

With masking \theAttacker cannot identify $auth$ token collisions: $aret_A$ and $aret_B$ have different mask values \pachash{0}{aret_A}
and \pachash{0}{aret_B}. Therefore it is impossible to identify a collision
with a
probability greater than by random selection.
This means that \theAttacker will succeed in the attack above with a
probability of $2^{-b}$.  We give a detailed proof in Appendix~\ref{sec:security-proofs}.

In practice, this means that \theAttacker can use this attack to traverse the
program's call graph, but cannot jump to an address that is not a valid return
address for function $C$.

\subsubsection{Violations that leave the call graph}\label{sec:evaluation-call-graph-escape}
We now consider \theAttacker's probability of success when attempting to return
to an address $ret_B$ in a way that that \emph{does not} follow the
program's call graph. (Summary in Table~\ref{tbl:evaluation-security}.)

In this case, the path from $B$ to $C$ has not been traversed, and the
instrumentation has never before computed the $auth$ token \pachash{ret_C}{aret_B}.
Therefore, \theAttacker succeeds at AG-Load---i.e., $\pachash{ret_C}{aret_B} = \pachash{ret_C}{aret_A}$---with
probability $\PP[\text{AG-Load}] = 2^{-b}$,
irrespective of whether the substituted \aret{B} is a valid authenticated return
address. On failure, which has probability $1-2^{-b}$, the process will crash.

\theAttacker's probability of then achieving goal AG-Jump depends on whether $ret_B$
is the return address of a valid call-site. If it is, then
\theAttacker can obtain a valid authenticated return pointer for that location
in
the same way as in Section~\ref{sec:evaluation-pac-collision}.
If \ret{B} has never been used
as a return address, then
no $auth$ token has ever been generated for that pointer and
AG-Jump is achieved with
probability at most $\PP[\text{AG-Jump}] = 2^{-b}$, independent of AG-Load.

\theAttacker can therefore succeed with probability $2^{-b}$ when
the return address is a valid call-site return address, or with probability of
$2^{-2b}$ when the return address is not.
\begin{table}
  \centering
  \begin{tabular}{|c|c|c|}
    \hline
    Violation type & No masking & Masking \\
    \hline
    On-graph & $1$ & $2^{-b}$ \\
    Off-graph to call-site & $2^{-b}$ & $2^{-b}$ \\
    Off-graph to arbitrary address & $2^{-2b}$ & $2^{-2b}$ \\
    \hline
  \end{tabular}
  \ifnotabridged\vspace{16pt}\fi
  \caption{
    Maximum success probability of call-stack integrity violations, with and without masking.
  }
  \label{tbl:evaluation-security}
\end{table}

\subsection{Run-time attack resistance of \IMPLNAME}
\label{sec:evaluation-security-runtime}

\IMPLNAME must ensure the integrity of \aret{n} and the confidentiality of the masks.
The former is achieved by storing \aret{n} in \gls{cr}, which is reserved for this purpose, not used by regular code, and hence, inaccessible to \theAttacker (Section~\ref{sec:function-call-instrumentation}).
The latter is maintained as the mask is re-generated each time it is needed and cleared after use (Section~\ref{sec:auth-token-masking}).
This holds true also in multi-threaded environments (Section~\ref{sec:implementation-multithreading}).
Traditional \gls{cfi} solutions are unable to withstand control-flow bending~\cite{Carlini15}: attacks where each control-flow transfer follows the program’s CFG, but the program execution trace conforms to no feasible benign execution trace.
Schemes like \IMPLNAME and shadow call stacks are not susceptible to backward-edge control-flow bending because they precisely protect the integrity of the return addresses.
\theAttacker cannot trick \IMPLNAME to deviate from an expected return flow by replacing \aret{n} with a valid, but outdated $aret$ value, because \IMPLNAME never writes \aret{n} onto the stack.
\theAttacker also cannot reliably exploit \gls{pac} collisions to replace part of the $aret$ chain, as each $aret$ is masked.
\theAttacker cannot tamper with the instrumentation itself by modifying the instructions in memory (Assumption~\ref{ass:dep}).
By requiring coarse-grained forward-edge \gls{cfi} (Assumption~\ref{ass:cfi}), \IMPLNAME ensures that $auth$ token calculations and masking are executed atomically and cannot be used to manipulate $ret_i$, $aret_{i-1}$ or the mask during the function prologue and epilogue.
This holds when the forward-edge \gls{cfi} is susceptible to control-flow bending (\Cref{sec:threat_model}).

\subsubsection{Tail calls and signing gadgets}
\label{sec:evaluation-security-tailcall}

A recent discovery by Google Project Zero~\cite{Aza19} shows that \gls{pa} schemes can be vulnerable to an attack whereby specific code sequences can be used as gadgets to generate \glspl{pac} for arbitrary pointers.
Recall that on \gls{pac} verification failure an \instr{aut} instruction removes the \gls{pac}, but corrupts a well-known high-order bit such that the pointer becomes invalid.
If a \instr{pac} instruction adds a \gls{pac} to a pointer $P$ with corrupt high-order bits, it treats the high-order bits \emph{as though they were correct} when calculating the new \gls{pac}, and flips a well-known bit $p$ of the \gls{pac} \emph{if} any high-order bit was corrupt.
This means that instruction sequences such as the one shown in Listing~\ref{lst:signing-gadget}, consisting of an \instr{aut} instruction followed by a \instr{pac} instruction, can be used generate a valid PAC for a pointer even if the original pointer is not valid to begin with. \theAttacker writes an arbitrary pointer $P$ to memory (\dOne) and allows it to be verified.
When verification fails, \instr{autia} removes the \gls{pac}, and corrupts the high-order bit in $P$, writing the resulting $P^*$ to the destination register (\dTwo).
The subsequent \instr{pacia} will add the \emph{correct \gls{pac} for $P$}, then flip bit $p$ of the \gls{pac} to indicate that the input pointer was invalid (\dThree).
\theAttacker can now flip bit $p$ back (\dFive) in order to obtain the correct \gls{pac} for pointer $P$ (\dSix).

\inputAsmListing{signing-gadget}{
  A \gls{pac} is based on the address bits.
An invalid input pointer (\dOne) after \instr{aut} (\dTwo) can be re-signed (\dThree), resulting in an output \gls{pac} with only a single bit-flip.
This could be exploited to generate valid \glspl{pac} for arbitrary pointers.
}

The \gls{pa} signing gadget requires finding a matching $\langle\autiplain, \paciplain\rangle$ pair operating on pointer $P$ in the code without any use of $P$ between these instructions.
In \IMPLNAME each verification is immediately followed by a return, which ensures that the failure is detected.
Tail calls are a notable exception.
Tail calls are function calls executed before return and optimized so that the callee directly returns to the caller of the optimized function.
For example, in Listing~\ref{lst:tail-call}, function \func{A} ends with a tail call to \func{B} using the \instr{b} instructions that does not update \gls{lr} (\dCOne).
The tail-called function can return (\dCTwo) to the \gls{lr} value set before the tail call (\dCThree).
\IMPLNAME limits \theAttacker to modifying the previous $auth$ token on the stack.
\theAttacker could attempt to exploit the signing gadget to trick \IMPLNAME to accept an invalid \aretAdv{i-1} (\dCFour), and subsequently load it into \gls{lr} after return.
However, \theAttacker cannot flip the bit $p$ of \aretAdv{i} (\dCFive) because \IMPLNAME guarantees it is immutable.
The invalid \aretAdv{i-1} is thus always passed into \autiplain (\dCFour) and so, detected at return from \func{B} (\dCTwo).
Forthcoming additions in the ARMv8.6-A architecture will preclude such attacks in general~\cite{ARMv86A}.

\inputAsmListing{tail-call}{
Tail calls on ARM replace the optimized call at the end of a function with a non-linking branch instruction (\dCOne).
}

\subsubsection{Sigreturn-oriented programming}
\label{sec:evaluation-sigreturn}

Sigreturn-oriented programming~\cite{Bosman14} is a exploitation technique in UNIX-like operating systems, including Linux, that abuses the \emph{signal frame} to take complete control of a process's execution state, i.e., the values of general purpose registers, \gls{sp}, \gls{pc}, status flags, etc. When the kernel delivers a signal, it suspends the process and changes the user-space processor context such that the appropriate signal handler is executed with the right arguments. When the signal handler returns, the original user-space processor context is restored. In a sigreturn attack \theAttacker sets up a fake signal frame and initiates a return from a signal that the kernel never delivered. Specifically, a program returns from the handler using a \texttt{sigreturn} system call that reads a signal frame (\texttt{struct sigcontext} in Linux) from the process stack.

Although a sigreturn attack is, in principle, problematic for \IMPLNAME (as it could allow \theAttacker control of any \texttt{EL0} register, including \gls{cr}), a number of defenses against sigreturn attacks have been proposed for the Linux kernel, any of which will protect \IMPLNAME{}.
Bosman and Bos~\cite{Bosman14} propose placing keyed signal canaries in the signal frame that are validated by the kernel before performing a \texttt{sigreturn}, or to keep a counter of the number of currently executing signal handlers. However, modern Linux versions rely solely on \gls{aslr}~\cite{Larsen14} to make it difficult for the attacker to trigger an unwarranted \texttt{sigreturn}. Fortunately \texttt{sigreturn} is never called directly from program code (in fact the GNU C library \texttt{sigreturn} simply returns an error value).
Instead the system call is triggered by signal trampoline code placed either in the kernel's \gls{vdso} or in the C library, both subject to \gls{aslr}.
For our chosen adversary model (Section \ref{sec:threat_model}) \gls{aslr} is not sufficient as \theAttacker can determine the contents of any readable memory in the process memory space.
However, \IMPLNAME itself, together with coarse-grained \gls{cfi} (Assumption \ref{ass:cfi}), ensures that \theAttacker cannot divert control flow from program code to the signal trampoline.
Nonetheless, 64-bit ARM programs that might call system calls directly using the \texttt{svc} instruction (without going through C library system call wrappers), would not be protected against the presence of such gadgets.
We discuss a potential general solution against sigreturn attacks that utilizes the \SHORTNAME construction in Appendix~\ref{sec:sigreturn-mitigation}.

\section{Performance Evaluation}
\label{sec:evaluation-performance}

At present, the only publicly available PA-enabled SoCs are the Apple A12, A13, S4, and S5, none of which support \gls{pa} for 3rd party code at the time of writing.
To verify the correctness of instrumentation we ran all benchmarks on the ARMv8-A \emph{Base Platform \gls{fvp}}, based on Fast Models 11.4, which supports ARMv8.3-A~\cite{ARMFVP}.
Because the \gls{fvp} runs the \version{4.14} kernel, we have used \gls{pa} RFC patches
\footnote{\url{https://lwn.net/Articles/752116/}}
modified to support all \gls{pa} keys.

The \gls{fvp} is not cycle-accurate and executes all instructions in one master cycle; therefore, it cannot be used for performance evaluation.
Based on prior evaluations of the QARMA cipher~\cite{Avanzi17}, which is used as the underlying cryptographic primitive in reference implementations of \gls{pa}~\cite{Qualcomm17}, Liljestrand \etal{} estimate that the \gls{pac} calculations incur an average overhead of four cycles on a 1.2GHz CPU~\cite{Liljestrand19}.
We employ the \emph{\glstext{pa}-analogue} \ifnotabridged(Listing~\ref{lst:pa-analogue})\fi introduced by Liljestrand \etal{} to estimate the run-time overhead of \IMPLNAME{}.
\ifnotabridged
The \glstext{pa}-analogue consists of four \instr{eor} instructions that both read and write the registers used by the corresponding \gls{pa} instruction in order to induce similar constraints on instruction pipelining within the CPU\@.
To preserve compiler behavior, the \glstext{pa}-analogue is swapped-in during a separate pre-emit pass, i.e., after both register allocation and instruction scheduling.

\inputAsmListing{pa-analogue}{
\glstext{pa}-analogue used to simulate overhead on non-PA hardware, based on an estimated overhead of 4 cycles.
Three exclusive-or inputs are constants, whereas the last instruction uses both inputs to ensure instruction pipelining must get both values.
}
\fi

\ifnotabridged

\todo{this needs to be updated for the non-abridged version}
Using the PA-analogue, we conducted benchmarks on a 96board Kirin 620 HiKey (LeMaker version) with an ARMv8-A Cortex A53 Octa-core CPU (1.2GHz) / 2GB LPDDR3 SDRAM (800MHz) / 8GB eMMC, running the Linux kernel $\mathrm{v}4.18.0$ and BusyBox $\mathrm{v}1.29.2$.
We have performed benchmarks using \ifnotabridged both nbench-byte-2.2.3\footnote{\url{http://www.math.utah.edu/~mayer/linux/bmark.html}} program and \fi the SPEC CPU 2017 benchmark package\footnote{\url{https://www.spec.org/cpu2017}}.

\subsection{nbench-byte-2.2.3}
\label{sec:nbench}

The nbench program includes 10 separate benchmarks and is designed to
measure CPU and memory performance. The benchmarks employ dynamic
workload adjustment to ensure that a test run takes at least a certain amount of
time. In order to determine the relative overhead introduced by \IMPLNAME, we took the same approach as prior work~\cite{Brasser17,Liljestrand19} and modified nbench to perform a pre-determined number of iterations of each benchmark and measured the execution time of each separately.
All binaries used in the performance evaluation were produced by our \IMPLNAME-enabled compiler.
We disabled all optimizations when compiling benchmark binaries (\texttt{-O0} flag for Clang and LLVM, and \texttt{-O=0} for \texttt{llc}).
We evaluated the performance of nbench in three configurations:
\begin{inparaenum}[i)]
\item \IMPLNAME disabled, to determine the baseline execution time;
\item \IMPLNAME enabled, without \gls{pac} masking; and
\item \IMPLNAME enabled, with \gls{pac} masking.
\end{inparaenum}
We repeated each benchmark 10 times and measured the user time using the \texttt{time} utility for each benchmark run.
The results are shown in Figure~\ref{fig:nbench_results}, and indicate an overhead of \overheadNbench{} when using \gls{pac} masking, and an overhead of $<\!0.3\%$ without (geometric mean of all benchmarks~\cite{Kouwe18}).

\begin{figure}[tp]
\centering
  \includegraphics[width=1\columnwidth]{nbench_results}
\caption{
Relative performance overhead of the individual nbench-byte-2.2.3 benchmarks.
The error bars indicate the standard error for $n=10$ test runs per benchmark.
The geometric mean of all benchmarks is \overheadNbench{}.
}
\label{fig:nbench_results}
\end{figure}
\fi

\subsection{SPEC CPU 2017}
\label{sec:spec}

\ifnotabridged{}
In contrast to nbench-byte-2.2.3, SPEC CPU 2017 is an industry-standard benchmarking suite that consists of larger units of work based on real-world applications.  
\fi
We ran benchmarks on Amazon EC2 using the SPEC CPU 2017 benchmark package\footnote{\url{https://www.spec.org/cpu2017}}.
To guarantee exclusive access to the hardware, we
used Amazon EC2 \texttt{a1.metal}\footnote{\url{https://aws.amazon.com/ec2/instance-types/a1/}}\ifaebadge\footnote{Performance evaluation on \texttt{a1.metal} instances was not part of the USENIX Security Artifact Evaluation process.}\fi instances, each with 16 64-bit ARMv8.2-A cores.
As these CPUs do not support \gls{pa}, we instrumented benchmarks with the \gls{pa}-analogue.
For comparison, we measured run-time overheads of:
\begin{inparaenum}[1)]
\item \shadowstack{} (a AArch64 production-ready software shadow call stack implementation for Clang 9~\cite{ShadowCallStack}),
\item \texttt{\branchProtFlag} (Clang's built-in \gls{pa}-based return address protection), and
\item \texttt{\stackProtFlag} (stack canaries).
\end{inparaenum}
We measured \IMPLNAME by instrumenting all function entry and exit points, excluding leaf functions that do not spill \gls{lr} or the \gls{cr} (this is similar to the heuristic used by \texttt{\branchProtFlag}).
We measured both full \IMPLNAME and \IMPLNAME without masking (\nomaskpacstack{}).

\shadowstack{} saves a function’s return address in a separately-allocated shadow stack and then uses the protected return address when performing a return.
On 64-bit ARM the \mReg{X18} register is reserved to hold a reference to the shadow stack.
\ifnotabridged{}
Current runtime support for \shadowstack{} is only available in Android's Bionic C library.
In addition, \shadowstack{} is only compatible with uninstrumented libraries which reserve the \mReg{X18} register, i.e., binaries built for a platform whose ABI reserves \mReg{x18},
\ifnotabridged{} (e.g., Android, Darwin, Fuchsia and Windows) \fi
or are compiled with the \texttt{-ffixed-x18} flag.
\fi
To perform a comparison against \IMPLNAME using the GNU C library (\texttt{glibc}) we ported \shadowstack{} support to \texttt{glibc}\ifabridged. \else
version 2.23 and compiled versions of our modified \texttt{glibc} and \texttt{libgcc} 6.4.1, the GCC low-level runtime library with the \texttt{-ffixed-x18} flag.
Our changes to \texttt{glibc} were based on revision da772e2\footnote{\url{https://android.googlesource.com/platform/bionic/+/da772e2113fad40575eea4ebbb522509be7dfe4f\%5E\%21/}} of the Bionic C library.
For \IMPLNAME measurements we used a prebuilt version of \texttt{glibc} 2.23 and \texttt{libgcc} 6.4.1 distributed by Linaro\footnote{\url{https://releases.linaro.org/components/toolchain/binaries/6.4-2018.05/}}.
\fi
Due to compatibility issues~\cite{Xu19}, we did not run the \texttt{perlbench} benchmarks with \shadowstack{}.

\begin{figure}[tp]
\centering
  \includegraphics[width=1\columnwidth]{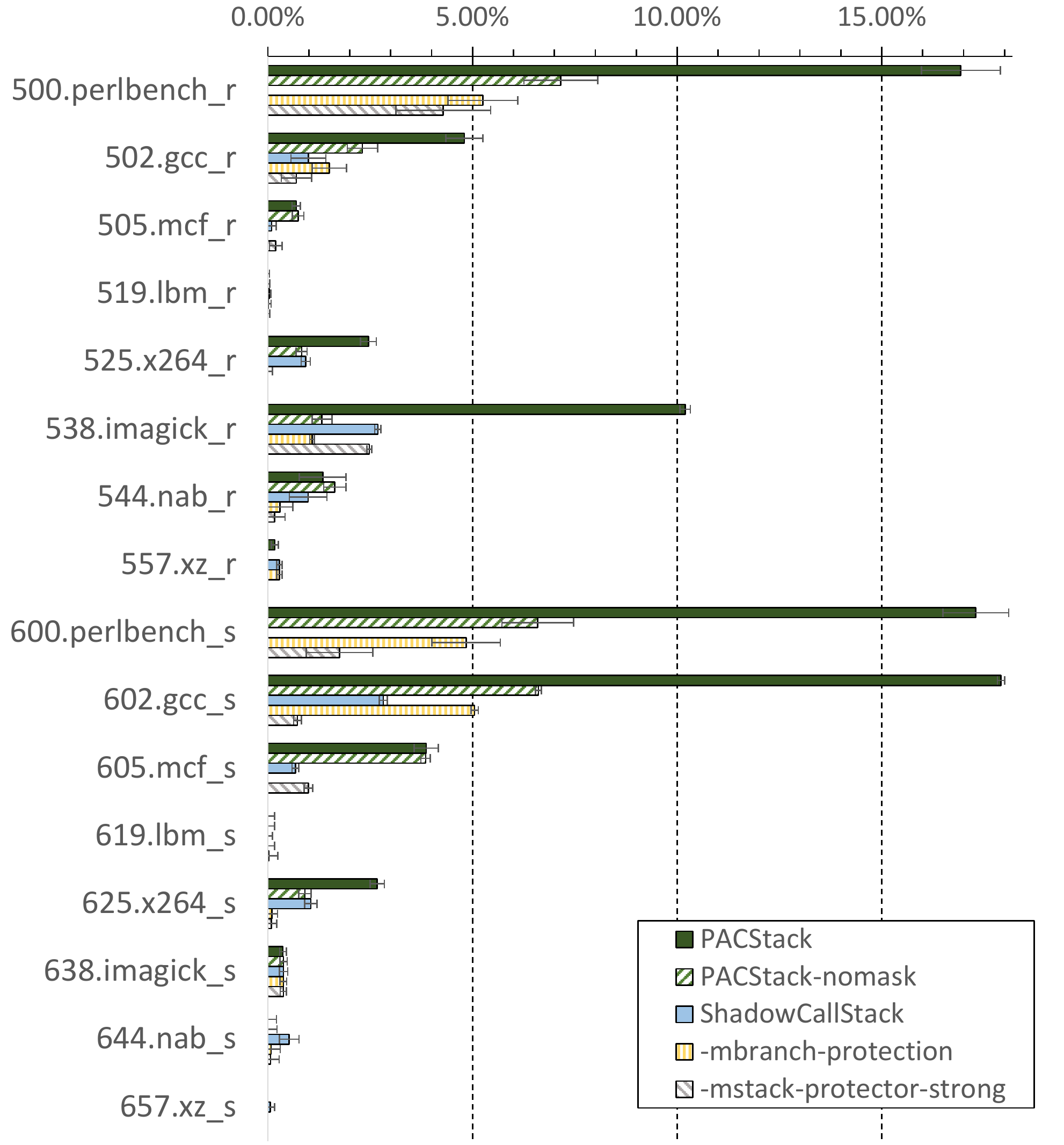}
\caption{
Relative performance overhead for SPEC CPU 2017 benchmarks as mean overhead over
baseline.  Error bars are 95\% confidence intervals.
}
\label{fig:spec_results}
\end{figure}

Our measurements include all C  SPECrate and SPECspeed benchmarks, compiled with \texttt{-O2} optimizations and flags to enable the measured instrumentation.
The suite is self-contained, avoiding the need to instrument system libraries.
For each benchmark, we compared the performance of the baseline (with all evaluated instrumentations disabled) to the measured configuration.
Figure~\ref{fig:spec_results} shows the mean overheads (w.r.t the baseline).
Table~\ref{tbl:spec-summary} shows the geometric mean of the
overheads, excluding \texttt{perlbench} which was incompatible with \shadowstack.
On C++ benchmarks we observed overheads of \overheadSCppNomask{} (\IMPLNAME)
and \overheadSCpp{} (\nomaskpacstack).
Due to compatibility issues with \shadowstack and \branchProtFlag, we limit our comparison to the C benchmarks.
\ifnotabridged
The SPEC CPU 2017 benchmark suite is resource intensive~\cite{Panda18}; a single iteration of all SPEC benchmarks in Figure~\ref{fig:spec_results} took 13 times longer than an iteration of all nbench benchmarks.
We therefore performed fewer measurements for SPEC than for nbench. Consequently, though the SPEC benchmarks are more representative of real-world workloads, they are more sensitive to outliers than those in Figure~\ref{fig:nbench_results}.
\fi

As expected, \texttt{\stackProtFlag} outperforms other instrumentations (but provides the weakest protection).
In terms of added instructions, \texttt{\branchProtFlag} is similar to \nomaskpacstack; the performance difference is likely due to \IMPLNAME reserving the \gls{cr} register and the additional store when saving it the stack.
\nomaskpacstack and \shadowstack{} have similar memory requirements (i.e., one extra store per function call), and show similar performance overheads.
The overhead of \IMPLNAME is proportional to the frequency of function
calls; benchmarks with few function calls are affected less than the benchmarks with frequent function calls.
For instance, the \texttt{519.lbm\_r} benchmark involves computations related to fluid dynamics and consists of large nested loops with few function calls.
Consequently we see little effect on the performance of \texttt{519.lbm\_r}.

Based on these results, we expect the overhead for both \IMPLNAME configurations to be
\begin{inparaenum}[a)]
  \item comparable to \shadowstack{}, and
  \item negligible on \ARMPA-capable hardware.
\end{inparaenum}

\begin{table}
\centering
\resizebox{0.9\columnwidth}{!}{
\begin{tabular}{l|c|c}
\toprule
                & SPECrate & SPECspeed \\
\midrule
\IMPLNAME       & 2.75\%   & 3.28\% \\
\nomaskpacstack & 0.86\%   & 1.56\% \\
\shadowstack    & 0.85\%   & 0.77\% \\
\branchProtFlag & 0.43\%   & 0.72\% \\
\stackProtFlag  & 0.43\%   & 0.25\% \\
\bottomrule
\end{tabular}
}\caption{
Geometric mean of measured overheads.
}\label{tbl:spec-summary}
\end{table}

\subsection{Real-world evaluation: \nginx}
\label{sec:evaluation-nxing}

We evaluated the efficacy of \IMPLNAME in a real-world setting using a \Gls{ssltps} test on the \nginx\footnote{\url{https://www.nginx.com/}} open source web server software.
\Gls{ssltps} measures a web server's capacity to create new SSL/TLS connections back to clients.
Clients send a series of HTTPS requests, each on a new connection.
The web server sends a 0‑byte response to each request.
The connection is closed after the response is received.
We chose the \gls{ssltps} test (instead of measuring throughput) to ensure that the load on the web server is CPU-bound, allowing us to estimate the upper bound for \IMPLNAME's impact on \nginx performance.

We conducted our tests on two separate Amazon EC2 A1 instances connected via elastic network interfaces with up to 10 Gbps capacity.
The web server (on an \texttt{a1.metal} instance, running  \nginx 1.17.8 with OpenSSL 1.1.1d) and the client (on an \texttt{a1.4xlarge} instance) ran the 64-bit ARM version of Ubuntu 18.04.
We configured the server to use the ECDHE-RSA-AES256-GCM-SHA384 cipher with a 2,048‑bit RSA key for HTTPS.
The client used \texttt{wrk}\footnote{\url {https://github.com/wg/wrk} (version of April 18, 2019)}, a modern HTTP benchmarking tool, to generate traffic.
We configured \texttt{wrk} in the same way as in a test on \nginx performance conducted by F5 Networks.\footnote{\url{https://www.nginx.com/blog/nginx-plus-sizing-guide-how-we-tested/}}
We ran a total of 15 copies of \texttt{wrk} on the client machine for 3 minutes each\ifabridged.\else and pinned each \texttt{wrk} process to different CPU on the client machine.
This resulted more consistent results compared to increasing the number of threads for a single \texttt{wrk} instance.\fi

We repeated the test with four and eight \nginx{} worker processes instrumented with \IMPLNAME and \nomaskpacstack, and compared the results with uninstrumented baseline performance.
In both configurations we also instrumented \nginx{}'s dependencies (OpenSSL, \texttt{pcre} and \texttt{zlib} libraries).
All binaries were compiled with \texttt{-O2} optimizations. We summarize the results in Table~\ref{tbl:nginx-performance}, showing
a 4--7\% overhead for \nomaskpacstack and 6--13\% overhead for \IMPLNAME.
These results are consistent with the performance overheads measures for SPEC CPU 2017 (\Cref{sec:spec}).

\begin{table}
\centering

\resizebox{\columnwidth}{!}{
\begin{tabular}{c|c|c|c|c|c|c|c|c}

\toprule
  \# of &
  \multicolumn{2}{c|}{Baseline} &
  \multicolumn{3}{c}{\nomaskpacstack} &
  \multicolumn{3}{c|}{\IMPLNAME} \\

  workers &
  req./sec. & $\sigma$ &
  req./sec. & $\sigma$ & overhead &
  req./sec. & $\sigma$ & overhead \\

\midrule
  4 & 14.2k & 142 &
      13.7k & 124 & 3.8\% &
      13.5k & 117 & 5.5\% \\
  8 & 30.7k & 722 &
      28.6k & 658 &  7.1\% &
      27.2k & 612 & 12.7\% \\

\bottomrule

\end{tabular}
}
\caption{
Requests/second, standard deviation ($\sigma$) and performance overhead for the \nginx{} \gls{ssltps} tests reported for both \IMPLNAME{} and \nomaskpacstack.
}
\label{tbl:nginx-performance}
\end{table}

\subsection{Compatibility testing using ConFIRM}
\label{sec:evaluation-confirm}

ConFIRM is a set small micro-benchmarking suite designed to test compatibility and relevance of \gls{cfi} solutions~\cite{Xu19}.
The suite is designed to test various corner-cases---e.g., function pointers, setjmp/longjmp and exception handling---that often cause compatibility issues for \gls{cfi} solutions.
ConFIRM is designed for x86-based architectures and includes some tests that are exclusive to the Microsoft Windows operating system.
Of the 18 64-bit Linux tests 11 compiled and worked on AArch64; these included virtual and indirect function calls, setjmp/longjmp, calling conventions, tail calls and load-time dynamic linking.
We ran these benchmarks on the \gls{fvp} (to guarantee functional equivalence to \gls{pa}-capable hardware) and confirmed that the tests passed with or without \IMPLNAME.

\section{Related Work}
\label{sec:related}

Control-flow hijacking have been known for more than two decades~\cite{Peslyak97}.  Most current \gls{cfi} solutions are \emph{stateless}: they validate each control-flow transfer in isolation without distinguishing among different paths in the \gls{cfg}.
\emph{Fully-precise static \gls{cfi}}~\cite{Carlini15} is the most restrictive stateless policy possible without breaking the intended functionality of the protected program.
In fully-precise static \gls{cfi}\ifnotabridged, and by extension any stateless policy,\else \ \fi the best possible policy for return instructions is to allow returns within a function $F$ to target any instruction that follows a call to $F$.
All stateless \gls{cfi} schemes, including fully-precise static \gls{cfi}, are
vulnerable to \emph{control-flow bending}~\cite{Carlini15}.

\emph{Stateful \gls{cfi}} can express policies that take previous control-flow transfers into account. \emph{HAFIX}~\cite{Davi15} is a hardware-assisted \gls{cfi} scheme that confines function returns to active call sites. Context-sensitive \gls{cfi}~\cite{vanderVeen15,Ding17,Hu18} further ensures that \emph{each} control-flow transfer taken by the program is consistent with a non-malicious trace.
Despite its better precision, context-sensitive \gls{cfi} enforcement is considered impractical for real-world adoption~\cite{Abadi09}. Hardware-assisted branch recording features available in modern 64-bit Intel microprocessors can be used to enable context-sensitive \gls{cfi} enforcement on commodity hardware, but suffer from
\begin{inparaenum}[i)]
\item limited branch history used to make \gls{cfi} decisions,
\item over-approximation of the program CFG, and
\item reliance on complex run-time monitoring.
\end{inparaenum}
\emph{HAFIX}, on the other hand, requires changes to the \ifnotabridged underlying processor architecture. \else processor. \fi

As dynamic schemes, \IMPLNAME and \emph{shadow call stacks}~\cite{Chiueh01,Frantzen01,Giffin02,Giffin04,Corliss05,Nebenzahl06,Abadi09,Dang15,Davi12,Tice14,Nyman17,Intel-CET,ShadowCallStack,Arnautov15} are not vulnerable to control-flow bending.
Stateless forward-edge \gls{cfi} enforcement is often combined with a shadow stack to enforce the integrity of return addresses stored on the call stack.
In fact, the results by Carlini \etal~\cite{Carlini15} show that a shadow stack (or equivalent mechanism) is essential for the security of \gls{cfi}.
\ifnotabridged
The shadow stack maintains a copy of each return address in a separate region of memory.
Each return instruction is then instrumented to validate that the return addresses on the call and shadow stack match.
This ensures that each return is restricted only to its corresponding call site.
\fi
However, traditional shadow stacks 
incur significant performance overhead and
lead to false positives for programming constructs that cause mismatches between calls and returns (C++ exceptions with stack unwinding, \setjmp /\longjmp).
Recent designs improve performance by either
leveraging a parallel shadow stack~\cite{Dang15}, or
using a dedicated register for shadow stack addressing~\cite{Burow19}.
But since the shadow stack in this schemes resides in the same address space as the target application, it can be compromised if \theAttacker knows its location.
A typical solution for dealing with mismatches between calls and returns is to pop return addresses off the shadow stack until a match is found, or the shadow stack is empty (e.g., binary RAD~\cite{Chiueh01}). This not only increases the complexity and run-time of the shadow stack instrumentation placed in the function epilogue, but also sacrifices precision, e.g., it allows \theAttacker to redirect \longjmp to any previously active call site. This can be avoided by storing and validating both the return address and stack pointer~\cite{Corliss05,Ozdoganoglu06,Tice14}.
So far, only hardware-assisted shadow stacks promise to achieve negligible overhead without security trade-offs (e.g., Intel CET\cite{Intel-CET}).

Park \etal~\cite{Park04} present a micro-architectural shadow stack implementation using the branch predictor \emph{return address stack}, a common hardware feature found in modern speculative superscalar processor designs.
The return address stack is typically a circular buffer; to avoid losing stored return addresses when the maximum capacity is reached, Park \etal{} modify the return address stack to spill a portion of its content to backup storage in main memory.
A Merkle-tree caching scheme is used to efficiently authenticate the backup storage before it is read back to the return address stack.
The latency of spill/fill operations on backup memory is
offset by the 100\% hit rate for branch prediction since
return addresses that exceed the return address stack capacity are retained.

The idea of using of MACs to protect the return address at run-time was introduced in \emph{Cryptographic \gls{cfi}}~(CCFI)~\cite{Mashtizadeh15} which uses MACs to protect return addresses and other control-flow data (e.g., function pointers and C++ vtable pointers). CCFI's return address protection is similar to \gls{pa}-based return address signing~\cite{Qualcomm17}; both bind the return address to the address of the function's stack frame and thus provide only coarse-grained resistance against pointer reuse~\cite{Liljestrand19}.
In contrast to \IMPLNAME, these approaches cannot prevent reuse attacks (See \Cref{sec:evaluation-security-reuse-in-practice}).
Independently to our work, Li \etal~\cite{Li19} propose a chain structure to protect return addresses but do not prevent the attacker from exploiting \gls{mac} collisions, and require custom hardware to realize their solution.

\emph{Program Counter Encoding}~\cite{Lee00,Cowan03,Frantzen01,Pyo02,Park17} protects return
addresses on the stack by encoding them with either a register-resident secret key~\cite{Lee00}, a read-only key stored in memory~\cite{Cowan03}, the \gls{sp}~\cite{Pyo02}, or the address at which the return address itself is stored (a.k.a.\ the \emph{self-address})~\cite{Park17}.
It is efficient, but relying on a secret key resident in user space makes such encoding schemes susceptible to buffer over-reads, and \gls{sp} or self-address encoding suffer the same drawbacks as \texttt{-msign-return-address}~\cite{Qualcomm17,Liljestrand19} (Section~\ref{sec:pa-return-address-protection}).

Other prominent defenses against control-flow attacks include fine-grained code randomization~\cite{Larsen14}, and code-pointer integrity (CPI)~\cite{Kuznetsov14}. Code randomization makes it more difficult for \theAttacker to find suitable gadgets to exploit,
but ineffective if \theAttacker knows the program memory layout. CPI protects code pointers by storing them in a separate \emph{safe stack}, which requires similar integrity guarantees as shadows stacks to remain effective~\cite{Evans15}.
Roessler \etal{} propose a metadata-tagged architecture to isolate stack-objects based on the stack-depth~\cite{Roessler18}.
However, similar to the \gls{sp} value (\Cref{sec:evaluation-security-reuse}), the stack-depth will repeat frequently during program execution.

\IMPLNAME targets the ARM architecture, which has received less attention compared to the x86 family of computer architectures in terms of \gls{cfi} research. \emph{MoCFI}~\cite{Davi12} is a software-based \gls{cfi} approach specifically targeting ARM application processors used in smartphones. It uses a combination of a shadow stack, static analysis and run-time heuristics to determine the set of valid targets for control-flow transfers, but suffers from the same drawbacks that plague traditional shadow stack schemes. \emph{CFI CaRE}~\cite{Nyman17} is a \gls{cfi} solution targeting small, embedded ARM-based microcontrollers (MCUs). It uses the ability to perform hardware-enforced isolated execution on ARMv8-M MCUs to isolate the shadow stack to a secure processor state. The ARMv8-M~\cite{ARMv8M} architecture enforces that calls to secure functions must target \emph{secure gate instructions} placed at the beginning of such functions. The ARMv8.5-A architecture introduces similar \emph{branch target indicators} (BTI)~\cite{ARMv8A} to ARM application processors. BTI constitutes one way to meet the \IMPLNAME pre-requisite of coarse-grained \gls{cfi} (\Cref{sec:threat_model}).

\section{Discussion}

\ifnotabridged
\subsection{Generalizing \SHORTNAME to other data structures}
\label{sec:discussion-generalizing-acs}

\SHORTNAME builds on the idea of chaining cryptographic authentication codes. This simple, yet powerful, construct is similar to hash chains, which have been used before as means of password protection (Lamport signatures~\cite{Lamport81}), digital signatures (Merkle trees~\cite{Merkle88}), and have seen use in technologies such as blockchain~\cite{Yaga18} and trusted hardware access control authorization policies~\cite{Arthur15}.

While the focus of this work is on applying this idea to protect the integrity
of return addresses in the program call stack, the same approach can be generalized to other data structures and applications.
For example, the call-stack protection could easily be extended to cover the \emph{frame pointer}, or other data stored in a function's stack frame, and protect such data from unauthorized modification. 

In addition to instrumentation that can protect the call stack, an ACS-like
authenticated stack, or other data structure such as a
Merkle-tree~\cite{Merkle88} can be implemented as reusable library, which would
allow application developers to protect the integrity of critical data structures from manipulation as a result of software~\cite{Chen05,Hu16}, or hardware attacks~\cite{Kim14}. 

An example of such a use case is data structures in operating system kernels.
For instance, the Linux kernel source code features a generic double linked list implementation, which doubles as a queue and stack, depending on where in the kernel it is used\footnote{\url{https://git.kernel.org/pub/scm/linux/kernel/git/torvalds/linux.git/tree/include/linux/list.h?h=v5.0}}.
Kernel data structures are critical to the system security.
Many of the vulnerabilities found in the kernel allow limited access to kernel data.
Malicious modification of kernel data can lead to a wide range of effects, including privilege escalation and process hiding~\cite{Azab14}.
Applying ACS-like protection to critical kernel stacks can protect such structures from:
\begin{inparaenum}[i)]
  \item malicious modification by \theAttacker in an effort to compromise kernel data integrity
  \item accidental misuse by programmers, e.g., operating on a stack as a queue and vice versa (a side-effect of reuse of generic list implementations).
\end{inparaenum}
\fi

\subsection{Support for software exceptions}
\label{sec:discussion-unwind}

The \setjmp~/ \longjmp interface has traditionally been used to provide exception-like functionality in C.
However, modern coding standards for C and C++ that aim to facilitate code safety, security, and reliability consider them harmful and forbid their use, e.g., MISRA C:2004~\cite[Rule 20.7]{Misra04} and JSF AV C++~\cite[Rule 20]{JsfAv05}.
Recall from Section~\ref{sec:design-unwind} that calling \longjmp with an expired \jmpbuf is undefined behavior.
For \IMPLNAME, this means that although the $aret_b$ in \jmpbuf is tied to the corresponding \mReg{SP} and $auth_i$, its \emph{freshness} cannot be guaranteed.
\theAttacker can modify \jmpbuf to contain the previously used $aret_b$ and $\mReg{SP}_b$, but must also modify the stack-frame at $\mReg{SP}_b$, such that it contains the prior $aret_{i}$.
This allows a control-flow transfer to a previously valid \setjmp return site and \mReg{SP} value.
To prevent reuse of expired \jmpbuf{} buffers, \longjmp{} can be rewound step-by-step, i.e., conceptually performing returns until the correct stack-frame is reached.

We plan to extend \IMPLNAME support to LLVM \libunwind
\footnote{\url{https://github.com/llvm/llvm-project/tree/master/libunwind}} --
it does frame-by-frame unwinding of the call stack.
By validating the \SHORTNAME on each stack frame unwinding, \IMPLNAME can ensure that a fresh and valid state is reached.

As C++ exceptions also cause irregular stack unwinding they pose a similar challenge.
But C++ does finer-grained stack unwinding to correctly destroy objects in unwound stack frames.
The LLVM \texttt{libcxxabi} library will, depending on configuration, use \libunwind for this purpose.
With \IMPLNAME support in \libunwind, we will be able to secure both \setjmp~/ \longjmp and support C++ exception handling.

\subsection{Interoperability with unprotected code}
\label{sec:discussion-legacy}

Interoperability with unprotected (uninstrumented) code is an
important deployment consideration. On one hand, \IMPLNAME-protected
applications may need to interoperate with unprotected shared
libraries. On the other, unprotected applications may need to
interoperate with \IMPLNAME-protected shared libraries. The latter
scenario is relevant for deployment in mobile operating systems like
Android, where multiple stakeholders provide application binaries to
consumer devices. The deployment of \IMPLNAME, or any other run-time
protection mechanism, is likely to be driven by OEMs that enable
specific protection schemes for the operating system and system
applications. However, OEMs are not in control of native code deployed
as part of applications.
It should be possible for one version of the shared libraries shipped with the operating system to remain interoperable with both \IMPLNAME-protected, and unprotected apps.

In Section~\ref{sec:function-call-instrumentation} we explain how the
use of callee-saved registers allows \IMPLNAME to remain interoperable
with unprotected code. Recall that because \gls{cr} is a callee-saved
register it will be restored upon return. However, \IMPLNAME cannot
guarantee that \gls{cr} remains unmodified during the execution of the
unprotected code that could temporarily store its value on the
stack. To meet the security guarantees
(Section~\ref{sec:evaluation-security}), \IMPLNAME instrumentation
must be applied to both the application and any shared libraries. But partial protection, e.g.\ \IMPLNAME-protected shared libraries can significantly raise the bar for the attacker, as calls into protected functions can still benefit from return address authentication.
Common shared libraries like \libc are a popular source for gadgets for run-time attacks because of their size and availability.
Because functions in a \IMPLNAME-protected library validate the return address in returns from library functions, they effectively remove a potentially large set of reusable gadgets from \theAttacker's disposal.

\section{Conclusion}
\label{sec:conclusion}

\SHORTNAME achieves security on-par with hardware-assisted shadow stacks (\Cref{sec:evaluation-security}).
With \IMPLNAME, we demonstrate how the general-purpose security \gls{pa} security mechanism can realize our design, \emph{without requiring additional hardware support or compromising security}.
Other general-purpose primitives like memory tagging and branch target indicators are being rolled out.
Creative uses of such primitives hold the promise of significantly improving software protection.

\ifnotanonymous{}
\section*{Acknowledgments}

This work was supported in part by NSERC (RGPIN-2020-04744), Intel
Collaborative Research Institute for Collaborative Autonomous \& Resilient
Systems (ICRI-CARS), and Google (ASPIRE program).
We acknowledge the computational resources provided by the Aalto Science-IT project.

\fi

\ifabridged{}
\bibliographystyle{plain}
\else
\bibliographystyle{ACM-Reference-Format}
\fi
{\normalsize\bibliography{references}}

\FloatBarrier
\appendix

\section{Security proofs}\label{sec:security-proofs}
In Section~\ref{sec:evaluation-security-reuse}, we gave an informal analysis of
the security of \SHORTNAME; here we give a more detailed proof of security, and
in particular prove that authentication token masking prevents \theAttacker from
obtaining exploitable authentication token collisions.

The argument proceeds as follows: we suppose that $\theAttacker$,
after obtaining $q$ authentication tokens, can find a pair of inputs $(x,y)$ and
$(x,y')$ whose authentication tokens \pachash{\cdot}{\cdot} collide. This can be used to
construct a distinguisher of the masks
$\pachash{0}{\cdot}$ from a random string. The structure of the authentication tags
is such that this further reduces to a semantic security game for one-time pad
encryption of the masks.
Then, we show that any violation of the integrity of an \SHORTNAME-protected
call stack also yields values whose authentication tokens collide as described
above, allowing us to bound the probability of an integrity violation.

We summarize our notation in Table~\ref{tbl:proofs-glossary}.
\begin{table}
  \centering
  \begin{tabular}{|p{0.20\linewidth}|p{0.20\linewidth}|p{0.5\linewidth}|}
    \hline
    \multicolumn{3}{|l|}{\textbf{Games}} \\
    \hline
    \small $G_\textsf{ACS}$\par \emph{\footnotesize (Figure~\ref{fig:acs-security-game})} & \multicolumn{2}{p{0.65\linewidth}|}{\footnotesize{}Security game for \SHORTNAME integrity.} \\
    \hline
    \small $G_\textsf{PAC-Collision}$  \emph{\footnotesize (Figure~\ref{fig:pac-collision})} & \multicolumn{2}{p{0.65\linewidth}|}{\footnotesize{}Security game for the identification of colliding authentication tokens.} \\
    \hline
    \small $G_\textsf{PAC-Distinguish}$ \emph{\footnotesize (Figure~\ref{fig:pac-distinguish})}  & \multicolumn{2}{p{0.65\linewidth}|}{\footnotesize{}Security game for the distinguishability of $\pachash{\cdot}{\cdot}$ from a random oracle.} \\
    \hline
    \small $G_1, G_2, G_3$ \emph{\footnotesize (Figure~\ref{fig:games})} & \multicolumn{2}{p{0.65\linewidth}|}{\footnotesize{}Semantic security games for the mask $\pachash{0}{\cdot}$.} \\
    \hline
    \hline
    \multicolumn{3}{|l|}{\textbf{Adversary interfaces}} \\
    \hline
    \small{}$G_\textsf{ACS}$ & \small{}$\theAttacker_\text{oracle-request}$ & \footnotesize{}Get path through the call-graph for which \theAttacker wants the final authenticated return address pushed to the stack. \\
                    & \small{}$\theAttacker_\text{oracle-response}$ & \footnotesize{}Return a previously-requested authenticated return address. \\
                    & \small{}$\theAttacker_\text{ACS-Violation}$ & \footnotesize{}Return to the challenger authenticated return values that can be used to violate call stack integrity. \\
    \hline
    \small{}$G_\textsf{PAC-Collision}$ & \small{}$\theAttacker_\text{oracle-request}$ & \footnotesize{}Get a value for which \theAttacker wants a masked authentication token.\\
                   & \small{}$\theAttacker_\text{oracle-response}$ & \footnotesize{}Return a previously-requested masked authentication token. \\
                    & \small{}$\theAttacker_\text{gen-collision}$ & \footnotesize{}Return to the challenger two authenticated return values with colliding authentication tokens. \\
    \hline
    \small{}$G_\textsf{PAC-Distinguish}$ & \small{}$\theAttacker_\text{oracle-request}$ & \footnotesize{}Get a value for which \theAttacker wants an authentication tag.\\
                   & \small{}$\theAttacker_\text{oracle-response}$ & \footnotesize{}Return a previously-requested authentication token.\\
                    & \small{}$\theAttacker_\text{distinguish}$ & \footnotesize{}Return to the challenger a single bit identifying whether the given tokens were from a random oracle or \pachash{\cdot}{\cdot}. \\
    \hline
    \small{}$G_1, G_2$ & \small{}$\mathcal{B}_\text{distinguish}$ & \footnotesize{}Identify the authentication token function used to generate masked authentication tokens. \\
    \hline
    \small{}$G_3$ & \small{}$\mathcal{B}_\text{distinguish'}$ & \footnotesize{}As for $G_1, G_2$, but with the inputs represented as strings, not functions.\\
    \hline
  \end{tabular}
  \ifnotabridged\vspace{12pt}\fi
  \caption{Notation used in Appendix~\ref{sec:security-proofs}.}
  \label{tbl:proofs-glossary}
\end{table}

\begin{figure}
  \centering
\fbox{
  \procedure[codesize=\scriptsize]{$G^\theAttacker_\textsf{PAC-Collision}(\secparam, H, q)$}{
    K \sample \bin^\secpar \\
    \pccomment{Give \theAttacker $q$ masked authentication tokens} \\ \pccomment{of their choice.} \\
    \pcfor i \in \{1, \ldots, q\} \pcdo \\
    \t (x,y) \gets \theAttacker_\text{oracle-request}() \\
    \t \theAttacker_\text{oracle-response}\left(\pachash{x}{y} \oplus
        \pachash{0}{y}\right) \\
    \pcendfor \\
    \pccomment{\theAttacker is challenged to provide inputs whose authentication
    tokens collide.} \\
    (\hat{x},\hat{y},\hat{y}') \gets \theAttacker_\text{gen-collision}() \\
    \pcif \hat{y} \ne \hat{y}' \wedge \pachash{\hat{x}}{\hat{y}} = \pachash{\hat{x}}{\hat{y}'} \pcthen \\
    \t \pcreturn 1 \\
    \pcendif \\
    \pcreturn 0
  }
}
\caption{Security game for finding colliding \glspl{pac} given masked
  authentication tokens.}
\label{fig:pac-collision}
\end{figure}

\begin{theorem}[\gls{pac}-masking prevents collision-finding]
  Suppose that after $q$ queries, an adversary \theAttacker can distinguish $\pachash{\cdot}{\cdot}$ from a
  random oracle with advantage no greater than
  $\adv^\theAttacker_\text{PAC-Distinguish}(\secparam,H,q)$, as given in
  Figure~\ref{fig:pac-distinguish}. Then, assuming a key-length of $\secpar$ for
  \pachash{\cdot}{\cdot}, and given access to
  $q$ masked authentication tokens, \theAttacker can identify a
  pair of inputs $(\hat{x},\hat{y})$ and $(\hat{x},\hat{y}')$ whose
  corresponding unmasked authentication tokens collide
  with advantage at most $2\adv^\theAttacker_\text{PAC-Distinguish}(\secparam,H,q)$.
\label{thm:masking}
\begin{proof}

We begin with a collision-game
$G^\theAttacker_\textsf{PAC-Collision}(\secparam,H,q)$, shown in Figure~\ref{fig:pac-collision} in which the adversary is
given oracle access to the authentication token generator and then asked to
provide values $x, y, y'$ such that $\pachash{x}{y} = \pachash{x}{y'}$.

\begin{figure}
  \centering
\fbox{
  \procedure[codesize=\scriptsize]{$G^\theAttacker_\textsf{PAC-Distinguish}(\secparam, H, q)$}{
    K \sample \bin^\secpar \\ \\
    \pccomment{$\mathcal{B}$ is given values of their choice from either} \\
    \pccomment{\pachash{\cdot}{\cdot} or a
      random oracle $RO(x,y)$} \\
    S_0(x,y) \stackrel{def}{=} RO(x,y) \\
    S_1(x,y) \stackrel{def}{=} \pachash{x}{y}\\ 
    c \sample \bin \\
    \pcfor i \in \{1, \ldots, q\} \pcdo \\
    \t (x,y) \gets \theAttacker_\text{oracle-request}() \\
    \t \theAttacker_\text{oracle-response}\left(S_c(x,y)\right) \\
    \pcendfor \\ \\
    \pccomment{\theAttacker is challenged to determine whether it received} \\ \pccomment{
      values from $\pachash{\cdot}{\cdot}$ or the random oracle.} \\
    \hat{c} \gets \theAttacker_\text{distinguish}() \\
    \pcif c \ne \hat{c} \pcthen \\
    \t \pcreturn 1 \\
    \pcendif \\
    \pcreturn 0
  }
}
\caption{Security game in which \theAttacker attempts to distinguish \pachash{\cdot}{\cdot} from a random oracle.}
\label{fig:pac-distinguish}
\end{figure}

An adversary that selects $(x,y,y')$ at random from $\bin^{\texttt{VA\_SIZE}} \times
\bin^{\texttt{VA\_SIZE} + b} \times \bin^{\texttt{VA\_SIZE} +
  b}$, such that $y \ne y'$, will
win with probability
$2^{-b}$; \theAttacker's advantage is therefore
\[
  \adv^\theAttacker_\textsf{PAC-Collision}(\secparam,H,q) =
  \PP\left[G^\theAttacker_\textsf{PAC-Collision}(\secparam,H,q) = 1\right] -
  2^{-b} .
\]

We will bound this advantage by reduction to a semantic security game for the
masks.  We consider the following games, shown in Figure~\ref{fig:games}, and
described in Figure~\ref{fig:game-hops}.

\begin{figure*}
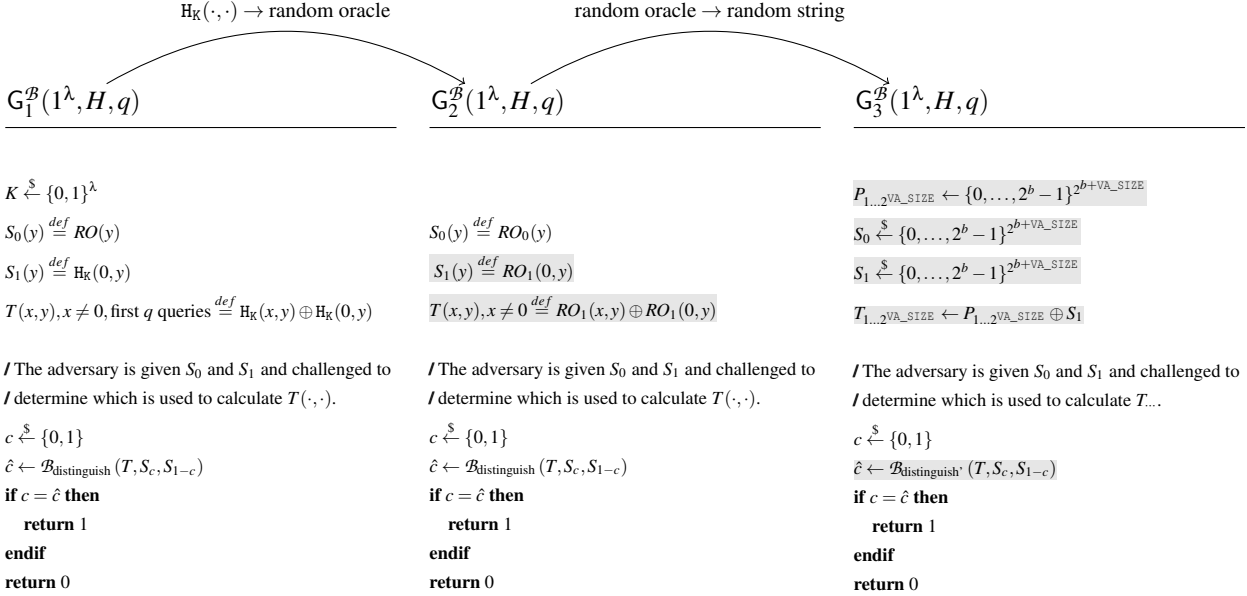

\vspace{34pt}
\begin{gameproof}[name=\mathsf{G}^\mathcal{B}, arg={(\secparam, H, q)}]
  \gameprocedure[codesize=\scriptsize]{
    \\
    K \sample \bin^\secpar\\
    S_0(y) \stackrel{def}{=} RO(y) \phantom{\bin^{2^b}}\\
    S_1(y) \stackrel{def}{=} \pachash{0}{y} \phantom{\bin^{2^b}}\\
    T(x,y), x \ne 0, \text{first $q$ queries} \stackrel{def}{=} \pachash{x}{y} \oplus \pachash{0}{y}
    \\ \\
    \pccomment{The adversary is given $S_0$ and $S_1$ and challenged to} \\
    \pccomment{determine which is used to calculate $T(\cdot,\cdot)$.} \\
    c \sample \bin \\
    \hat{c} \gets \mathcal{B}_\text{distinguish}\left(T, S_c, S_{1-c}\right) \\
    \pcif c = \hat{c} \pcthen \\
    \t \pcreturn 1 \\
    \pcendif \\
    \pcreturn 0
  }
  \pchspace
  \gameprocedure[codesize=\scriptsize]{
    \\ \phantom{\bin^{2^b}}\\
    S_0(y) \stackrel{def}{=} RO_0(y) \\
    \gamechange{
    \gamechange{$S_1(y) \stackrel{def}{=} RO_1(0,y)$}} \phantom{\bin^{2^b}}\\
    \gamechange{$T(x,y), x \ne 0  \stackrel{def}{=} RO_1(x,y) \oplus
        RO_1(0,y)$} \\ \\
    \pccomment{The adversary is given $S_0$ and $S_1$ and challenged to} \\
    \pccomment{determine which is used to calculate $T(\cdot,\cdot)$.} \\
    c \sample \bin \\
    \hat{c} \gets \mathcal{B}_\text{distinguish}\left(T, S_c, S_{1-c}\right) \\
    \pcif c = \hat{c} \pcthen \\
    \t \pcreturn 1 \\
    \pcendif \\
    \pcreturn 0
  }
  \pchspace
  \gameprocedure[codesize=\scriptsize]{
    \phantom{\bin^{2^b}}\\
    \gamechange{$P_{1\ldots2^{\texttt{VA\_SIZE}}} \gets \{0, \ldots, 2^b - 1\}^{2^{b + \texttt{VA\_SIZE}}}$} \\
    \gamechange{$S_0 \sample \{0, \ldots, 2^b - 1\}^{2^{b + \texttt{VA\_SIZE}}}$} \phantom{\stackrel{def}{=}}\\
    \gamechange{$S_1 \sample \{0, \ldots, 2^b - 1\}^{2^{b + \texttt{VA\_SIZE}}}$} \phantom{\stackrel{def}{=}} \\
    \gamechange{$T_{1\ldots2^{\texttt{VA\_SIZE}}} \gets P_{1\ldots2^{\texttt{VA\_SIZE}}} \oplus
        S_1 $} \phantom{\stackrel{def}{=}}\\ \\
    \pccomment{The adversary is given $S_0$ and $S_1$ and challenged to} \\
    \pccomment{determine which is used to calculate $T_{\cdots}$.} \\
    c \sample \bin \\
    \gamechange{$\hat{c} \gets \mathcal{B}_\text{distinguish'}\left(T, S_c, S_{1-c}\right)$} \\
    \pcif c = \hat{c} \pcthen \\
    \t \pcreturn 1 \\
    \pcendif \\
    \pcreturn 0
  }
  \addgamehop{1}{2}{hint={\footnotesize $\pachash{\cdot}{\cdot} \rightarrow \text{random oracle}$}}
  \addgamehop{2}{3}{hint={\footnotesize $\text{random oracle} \rightarrow \text{random string}$}}
\end{gameproof}
\caption{Security games used in Theorem~\ref{thm:masking}.}\label{fig:games}
\end{figure*}

\begin{figure}
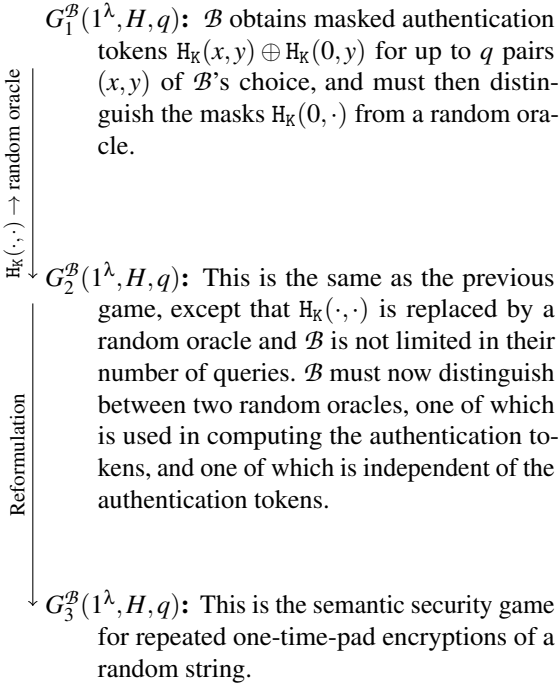

  \centering
  \begin{minipage}[c]{0.8\linewidth}
\begin{gamedescription}[name={G^\mathcal{B}},arg={(\secparam, H, q)}]
  \describegame $\mathcal{B}$ obtains masked authentication tokens
  $\pachash{x}{y} \oplus \pachash{0}{y}$ for up to $q$ pairs $(x, y)$
  of $\mathcal{B}$'s choice, and must then distinguish the masks
  $\pachash{0}{\cdot}$ from a random oracle.

  \vspace{3em}

  \describegame[length=8em,inhint={\footnotesize $\pachash{\cdot}{\cdot} \rightarrow \text{random
      oracle}$}] This is the same as the previous game, except that
  $\pachash{\cdot}{\cdot}$ is replaced by a random oracle and $\mathcal{B}$ is not
  limited in their number of queries. $\mathcal{B}$ must now distinguish between
  two random oracles, one of which is used in computing the authentication
  tokens, and one of which is independent of the authentication tokens.

  \vspace{2em}

  \describegame[length=11.5em,inhint={\footnotesize Reformulation}]
  This is the semantic security game for repeated one-time-pad encryptions of a
  random string.
\end{gamedescription}
\end{minipage}
\caption{The game-hops used in Figure~\ref{fig:games}.}
\label{fig:game-hops}
\end{figure}

The first hop, from $\mathsf{G_1}$ to $\mathsf{G_2}$, is based on
indistinguishability and relaxation: we suppose that $\pachash{\cdot}{\cdot}$ can be distinguished from
a random oracle with probability no more than $\frac{1}{2} +
\adv^\theAttacker_\textsf{PAC-Distinguish}(\secparam, H, q)$, and that the
adversary is not limited in the number of queries that can be made to the masked
authentication token oracle.  Then,
\begin{align*}
  \PP[G_1^\mathcal{B}(\secparam, H, q) = 1] \le\quad &\PP[G_2^\theAttacker(\secparam,
  H, q) = 1] \\ &+
  \adv^\theAttacker_\textsf{PAC-Distinguish}(\secparam, H, q) .
\end{align*}
The second hop, from $\mathsf{G_2}$ to $\mathsf{G_3}$, is a mere reformulation
of $\mathsf{G_2}$ such that random oracles are represented as strings, and
that rather than allowing $\mathcal{B}$ to request arbitrarily many
authentication tokens from the challenger, we instead give $\mathcal{B}$ direct
access to the oracle, as represented by the sequence of strings
$T_{1\ldots2^\texttt{VA\_SIZE}}$.

The third game is a semantic security game for the one-time pad, where
\theAttacker is given $2^\texttt{VA\_SIZE}$ encryptions of $S_1$ and then asked to
distinguish between $S_1$ and a random string.  The perfect secrecy of the
one-time pad means that $\PP[G_1^\mathcal{B}(\secparam) = 1] = \frac{1}{2}$ and
so
\begin{align}
  \PP[G_1^\mathcal{B}(\secparam) = 1] \le \frac{1}{2} +
  \adv^\theAttacker_\textsf{PAC-Distinguish}(\secparam,H,q) . \label{eqn:g1bound}
\end{align}
Finally, we provide a reduction from
$\mathsf{G}_\textsf{PAC-Collision}^\theAttacker(\secparam,H,q)$ to
$\mathsf{G_1}^\mathcal{B}(\secparam)$. Suppose \theAttacker can win
$\mathsf{G}_\textsf{PAC-Collision}^\theAttacker(\secparam,H,q)$ with advantage
$\adv^\theAttacker_\text{PAC-Collision}(\secparam,H,q)$. Then, we define an adversary
$\mathcal{A}^\theAttacker$ for $\mathsf{G_1}^\mathcal{B}(\secparam)$, shown in
Figure~\ref{fig:adversary-distinguish}.

\begin{figure}
  \centering
  \begin{minipage}[c]{0.58\linewidth}
    \fbox{\vbox{
        \procedure[codesize=\scriptsize]{$\mathcal{B}^\theAttacker_\text{oracle-request}()$}{
          \pcreturn \theAttacker_\text{oracle-request}() \\
          \phantom{\pcif S(y) \oplus S(y') = T(x,y) \oplus T(x,y') \pcthen}
        }

        \procedure[codesize=\scriptsize]{$\mathcal{B}^\theAttacker_\text{oracle-response}(x)$}{
          \theAttacker_\text{oracle-response}(x) \\
          \phantom{\pcif S(y) \oplus S(y') = T(x,y) \oplus T(x,y') \pcthen}
        }

        \procedure[codesize=\scriptsize]{$\mathcal{B}^\theAttacker_\text{distinguish}(T, S, S')$}{
          x, y, y' \gets \theAttacker_\text{gen-collision}(T) \\
          \pcif S(y) \oplus S(y') = T(x,y) \oplus T(x,y') \pcthen \\
          \t \pcreturn 1 \\
          \pcelse \\
          \t \pcreturn 0 \\
          \pcendif
        }
      }}
  \end{minipage}
  
  \caption{An adversary $\mathcal{B}^\theAttacker$ for $G_1$ used in our black-box reduction of
    $G_\textsf{PAC-Collision}$ to $G_1$.  Not shown is the variant
    $\mathcal{B}^\theAttacker_\text{distinguish'}(T, S, S')$ that is identical to
    $\mathcal{B}^\theAttacker_\text{distinguish}(T, S, S')$ except that $T$, $S$,
    and $S'$ are given in the form of strings.}
  \label{fig:adversary-distinguish}
\end{figure}

This adversary wins $\mathsf{G_1}^\mathcal{B}(\secparam)$ with probability at
least $\frac{1}{2} +
\frac{1}{2}\adv^\theAttacker_\text{PAC-Collision}(\secparam, H, q)$, and so
by (\ref{eqn:g1bound})
\[
\adv^\theAttacker_\text{PAC-Collision}(\secparam, H, q) \le 2
\adv^\theAttacker_\text{PAC-Distinguish}(\secparam,H,q) .
\]
If the \gls{mac} $\pachash{\cdot}{\cdot}$ is a pseudo-random function family with respect to
$K$, then $\adv^\theAttacker_\text{PAC-Distinguish}(\secparam,H,q)$ is negligible, and thus so is
$\adv^\theAttacker_\text{PAC-Collision}(\secparam,H,q)$.
\end{proof}
\end{theorem}

With a bound on \theAttacker's probability of successfully obtaining a \gls{pac}
collision, we may now obtain a bound on their probability of violating the
integrity of an \SHORTNAME-protected call stack.

\begin{theorem}[Security of \SHORTNAME]
  Consider a program whose call stack is protected by \SHORTNAME, which has a
  call-graph $C$ and $b$-bit masked authentication tokens $\pactag{x}{y} =
  \pachash{x}{y} \oplus \pacmask{y}$. Then, an adversary with
  arbitrary control over memory can violate backward-edge control-flow integrity
  with probability
  \begin{align*}
    \PP\left[G^\theAttacker_\textsf{\SHORTNAME}(\secparam,H,C,q)\right]
    &\le \PP\left[G^\theAttacker_\textsf{PAC-Collision}(\secparam,H,q)\right] \\
    &\le 2^{-b} + 2 \adv^\theAttacker_\text{PAC-Distinguish}(\secparam, H, q)
  \end{align*}
  \begin{proof}
    We begin with a security game for \SHORTNAME, shown in
    Figure~\ref{fig:acs-security-game}.
    
    Our goal is to provide a black-box reduction from
    $G^\theAttacker_\textsf{\SHORTNAME}(\secparam, H, C, q)$ to
    $G^\theAttacker_\textsf{PAC-Collision}(\secparam, H, q)$.

    From
    line~24 of Figure~\ref{fig:acs-security-game}, winning
    $G^\theAttacker_\textsf{\SHORTNAME}$ implies that \theAttacker has obtained
    colliding authentication tokens, and therefore \theAttacker can win
    $G^\theAttacker_\textsf{PAC-Collision}$ with probability at least $\PP[G^\theAttacker_\textsf{\SHORTNAME}]$.
    Substituting the bound from Theorem~\ref{thm:masking}, we obtain the bound given.
  \end{proof}
\end{theorem}

\section{Mitigation of sigreturn attacks}\label{sec:sigreturn-mitigation}

A solution for precluding sigreturn attacks against \IMPLNAME would be to include the \emph{signal return value} to the \IMPLNAME chain via the \gls{pc} value stored on the signal frame:
\begin{align*}
  \asigret{i} =
\begin{cases}
    \pachash{\sigret{i}}{ \asigret{i-1}}  & \quad \text{if } i > 0\\
    \pachash{\sigret{i}}{ \aret{n}}  & \quad \text{if } i = 0\\
  \end{cases}
\end{align*}
Upon signal delivery, the kernel stores a copy of \asigret{n} securely in kernel space as a reference value.
If the process was already executing a signal handler, and thus the kernel already has a reference copy of \asigret{n-1} on record, it stores \asigret{n-1} in the new signal frame and overwrites the secure copy with \asigret{n}.
On \texttt{sigreturn} the kernel attempts to validate the \gls{pc} and \gls{cr} values in the signal frame as though the reference value was \asigret{0}.
If successful it performs the signal return to \sigret{n} and restores \aret{n} to \gls{cr}.
Otherwise the kernel assumes a return to a nested signal handler, and retrieves \sigretAdv{n} and \asigretAdv{n-1} from the signal frame, validates them by calculating \asigretAdv{n} = \pachash{\sigretAdv{n}}{\asigretAdv{n-1}} and comparing the result against the stored \asigret{n} reference value.
If successful the kernel replaces \asigret{n} with \asigret{n-1} in the secure kernel store and performs the signal return to \sigret{n}.
If the validation fails the kernel terminates the process.
This prevents \theAttacker from 
\begin{inparaenum}[1)]
 \item overwriting \gls{cr}, and
 \item forging the \gls{pc} values in signal frames.
\end{inparaenum}
For general protection against sigreturn attacks corrupting any register stored in the signal frame, all register values could be included in the \asigret{} calculation using the \texttt{pacga} instruction and validated at the time of \texttt{sigreturn}. 

\begin{figure}[t]
  \centering
  \fbox{
    \procedure[codesize=\scriptsize,linenumbering]{$G^\theAttacker_\textsf{\SHORTNAME}(\secparam, H, C, q)$}{
      K \sample \bin^\secpar \\
      \pccomment{Give \theAttacker $q$ tokens from call-graph traversals.} \\
      \pcfor i \in \{1, \ldots, q\} \pcdo \\
      \t p_{1\ldots m+1} \gets \theAttacker_\textrm{oracle-request}() \\
      \t \pccomment{Is the request for a real path through the call-graph?} \\
      \t \pcif \exists j: p_j \rightarrow p_{j+1} \notin \mathrm{edges}(C) \pcthen \\
      \t \t \pcreturn 0 \\
      \t \pcendif \\
      \t auth_m \gets \pactag{p_{m}}{\pactag{p_{m-1}}{\cdots} \parallel p_{m-1}} \parallel p_{m} \\
      \t \theAttacker_\textrm{oracle-response}(auth_m) \\
      \pcendfor \\
      ptr_\textsf{jumper}, ptr_\textsf{correct}, auth_\textsf{correct},
      t_\textsf{correct}, \\ \t \t ptr_\textsf{adv}, auth_\textsf{adv},
      t_\textsf{adv} \gets \theAttacker_\textsf{\SHORTNAME-Violation}() \\
      \pccomment {The substituted masked authenticated return address must be different.} \\
      \pcif ptr_\textsf{correct} = ptr_\textsf{adv} \wedge auth_\textsf{correct} = auth_\textsf{adv} \pcthen \\
      \t \pcreturn 0 \\
      \pcendif \\ 
      \pccomment{Does the return pointer authenticate correctly with the
      adversary's} \\ \pccomment{new masked authenticated return address as the modifier?} \\
      \pcif \pachash{ptr_\textsf{jumper}}{auth_\textsf{correct} \parallel
      ptr_\textsf{correct}}\\ \nonumber \t \ne
      \pachash{ptr_\textsf{jumper}}{auth_\textsf{adv} \parallel
      ptr_\textsf{adv}}\pcthen \\
      \t \pcreturn 0  \\
      \pcendif \\ 
      \pccomment{Did the adversary provide a valid masked authenticated return address?} \\
      \pcif auth_\textsf{adv} = \pachash{ptr_\textsf{adv}}{t_\textsf{adv}} \\
      \t \pcreturn 1 \\
      \pcendif \\
      \pcreturn 0
      }
    }
    \caption{Security game for \SHORTNAME with respect to a program having
      call-graph $C$ and authentication token function $\pactag{\cdot}{\cdot}$.}
    \label{fig:acs-security-game}
  \end{figure}

\ifnotabridged
\twocolumn
\section{Glossary}
\begin{adjustbox}{width=\textwidth}
\begin{tabular}{|r|l|}
  \hline
  Return address & An address that is used as the target of a return instruction \\
  \hline
  Authenticated pointer & A pointer containing some embedded data that can be used to validate its authenticity \\
  \hline
  Authenticated return address & An authenticated pointer whose `pointer' part is a return address. \\
  \hline
  PAC & The value $\pachash{\cdot}{\cdot}$ produced by ARM-\gls{pa}. \\
  \hline
  Mask & The pseudo-random value \pachash{0}{\aret{i-1}} \\
  \hline
  Authentication token & $\pachash{ret_i}{\pachash{ret_{i-1}}{\cdots} \parallel ret_{i-1}}$ in \SHORTNAME;$\pachash{ret_{i}}{\mReg{SP}}$ in~\cite{Qualcomm17} \\
  \hline
  Masked authentication token & The authentication token exclusive-OR-ed with the mask \\
  \hline
\end{tabular}
\end{adjustbox}
\begin{table*}
  \section{\ARMPAFULL Instructions\ifnotabridged\vspace{12pt}\fi}\label{sec:appendix-instructions}
\centering
\resizebox{0.6\textwidth}{!}{
\begin{tabular}{c|l|c|c|c|c|c|c|c|c}
\toprule
  \multirow{3}{*}{Instruction} &
  \multirow{3}{*}{Mnemonic} &
  \multicolumn{5}{c|}{PA Key} &
  \multirow{3}{*}{Addr.} &
  \multirow{3}{*}{Mod.} &
  \multirow{3}{*}{\parbox{1.5cm}{Backwards-compatible}} \\
  & & \multicolumn{2}{c|}{Instr.} & \multicolumn{2}{c|}{Data} & Gen-& & & \\
  & & A & B & A & B & eric & & & \\
\midrule
  \multicolumn{10}{c}{BASIC POINTER AUTHENTICATION INSTRUCTIONS} \\
\midrule
  \multirow{10}{*}{\parbox{3.5cm}{Add PAC to instr. addr.}}
  & \instr{paciasp}   & \tableYes{} & & & & & LR          & SP            & \tableYes{} \\
  & \instr{pacia}     & \tableYes{} & & & & & \textit{Xd} & \textit{Xm}   & \tableNo{} \\
  & \instr{paciaz}    & \tableYes{} & & & & & LR          & \textit{zero} & \tableYes{} \\
  & \instr{paciza}    & \tableYes{} & & & & & \textit{Xd} & \textit{zero} & \tableNo{} \\
  & \instr{pacia1716} & \tableYes{} & & & & & X17         & X16           & \tableYes{} \\
  & \instr{pacibsp}   & & \tableYes{} & & & & LR          & SP            & \tableYes{} \\
  & \instr{pacib}     & & \tableYes{} & & & & \textit{Xd} & \textit{Xm}   & \tableNo{} \\
  & \instr{pacibz}    & & \tableYes{} & & & & LR          & \textit{zero} & \tableYes{} \\
  & \instr{pacizb}    & & \tableYes{} & & & & \textit{Xd} & \textit{zero} & \tableNo{} \\
  & \instr{pacib1716} & & \tableYes{} & & & & X17         & X16           & \tableYes{} \\
\midrule
  \multirow{4}{*}{\parbox{3.5cm}{Add PAC to data addr.}}
  & \instr{pacda}    & & & \tableYes{} & & & \textit{Xd} & \textit{Xm,}  & \tableNo{} \\
  & \instr{pacdza}   & & & \tableYes{} & & & \textit{Xd} & \textit{zero} & \tableNo{} \\
  & \instr{pacdb}    & & & & \tableYes{} & & \textit{Xd} & \textit{Xm}   & \tableNo{} \\
  & \instr{pacdzb}   & & & & \tableYes{} & & \textit{Xd} & \textit{zero} & \tableNo{} \\
\midrule
  \multirow{1}{*}{\parbox{3.5cm}{Calculate generic MAC}}
  & \instr{pacga} & & & & & \tableYes{} & & & \tableNo{} \\
\midrule
  \multirow{10}{*}{\parbox{3.5cm}{Authenticate instr. addr.}}
  & \instr{autiasp}   & \tableYes{} & & & & & LR          & SP            & \tableYes{} \\
  & \instr{autia}     & \tableYes{} & & & & & \textit{Xd} & \textit{Xm}   & \tableNo{} \\
  & \instr{autiaz}    & \tableYes{} & & & & & LR          & \textit{zero} & \tableYes{} \\
  & \instr{autiza}    & \tableYes{} & & & & & \textit{Xd} & \textit{zero} & \tableNo{} \\
  & \instr{autia1716} & \tableYes{} & & & & & X17         & X16           & \tableYes{} \\
  & \instr{autibsp}   & & \tableYes{} & & & & LR          & SP            & \tableYes{} \\
  & \instr{autib}     & & \tableYes{} & & & & \textit{Xd} & \textit{Xm}   & \tableNo{} \\
  & \instr{autibz}    & & \tableYes{} & & & & LR          & \textit{zero} & \tableYes{} \\
  & \instr{autizb}    & & \tableYes{} & & & & \textit{Xd} & \textit{zero} & \tableNo{} \\
  & \instr{autib1716} & & \tableYes{} & & & & X17         & X16           & \tableYes{} \\
\midrule
  \multirow{4}{*}{\parbox{3.5cm}{Authenticate data addr.}}
  & \instr{autda}    & & & \tableYes{} & & & \textit{Xd} & \textit{Xm}  & \tableNo{} \\
  & \instr{autdza}   & & & \tableYes{} & & & \textit{Xd} & \textit{zero} & \tableNo{} \\
  & \instr{autdb}    & & & & \tableYes{} & & \textit{Xd} & \textit{Xm}   & \tableNo{} \\
  & \instr{autdzb}   & & & & \tableYes{} & & \textit{Xd} & \textit{zero} & \tableNo{} \\
\midrule
  \multirow{3}{*}{\parbox{3.5cm}{Strip PAC}}
  & \instr{xpacd}   & & & & & & \textit{Xd} & & \tableNo{} \\
  & \instr{xpaci}   & & & & & & \textit{Xd} & & \tableNo{} \\
  & \instr{xpaclri} & & & & & & LR          & & \tableYes{} \\
\midrule
  \multicolumn{10}{c}{COMBINED POINTER AUTHENTICATION INSTRUCTIONS} \\
\midrule
  \multirow{2}{*}{\parbox{3.5cm}{Authenticate instr. addr. and return}} &
    \instr{retaa}    & \tableYes{} & & & & & LR & SP & \tableNo{} \\
  & \instr{retab}    & & \tableYes{} & & & & LR & SP & \tableNo{} \\
\midrule
  \multirow{4}{*}{\parbox{3.5cm}{Authenticate instr. addr. and branch}}
  & \instr{braa}     & \tableYes{} & & & & & \textit{Xd} & \textit{Xm}   & \tableNo{} \\
  & \instr{braaz}    & \tableYes{} & & & & & \textit{Xd} & \textit{zero} & \tableNo{} \\
  & \instr{brab}     & & \tableYes{} & & & & \textit{Xd} & \textit{Xm}   & \tableNo{} \\
  & \instr{brabz}    & & \tableYes{} & & & & \textit{Xd} & \textit{zero} & \tableNo{} \\
\midrule
  \multirow{4}{*}{\parbox{3.5cm}{Authenticate instr. addr. and branch with link}}
  & \instr{blraa}    & \tableYes{} & & & & & \textit{Xd} & \textit{Xm}   & \tableNo{} \\
  & \instr{blraaz}   & \tableYes{} & & & & & \textit{Xd} & \textit{zero} & \tableNo{} \\
  & \instr{blrab}    & & \tableYes{} & & & & \textit{Xd} & \textit{Xm}   & \tableNo{} \\
  & \instr{blrabz}   & & \tableYes{} & & & & \textit{Xd} & \textit{zero} & \tableNo{} \\
\midrule
  \multirow{2}{*}{\parbox{3.5cm}{Authenticate instr. addr. and exception return}}
  & \instr{eretaa}   & \tableYes{} & & & & & ELR & SP & \tableNo{} \\
  & \instr{eretab}   & & \tableYes{} & & & & ELR & SP & \tableNo{} \\
\midrule
  \multirow{2}{*}{\parbox{3.5cm}{Authenticate data. addr. and load register}}
  & \instr{ldraa}    &  & & & \tableYes{} & & \textit{Xd} & \textit{zero} & \tableNo{} \\
  & \instr{ldrab}    & &  & & & \tableYes{} & \textit{Xd} & \textit{zero} & \tableNo{} \\
\bottomrule
\end{tabular}
}
\ifnotabridged\vspace{16pt}\fi
  \caption{
  List of \gls{pa} instructions~\cite{Liljestrand19}.
  \emph{PA Key} indicates the PA key the instruction uses.
  \emph{Addr.} indicates the source of the address to be signed or authenticated.
  \emph{Mod.} indicates the modifier used by the instruction.
  \textit{Xd} and \textit{Xm} indicates that the input is taken from a general purpose register. 
  The \emph{backwards-compatible} column indicates if the instruction is safe on pre ARMv8.3-A.}
\label{tab:pa-instructions}
\end{table*}
\fi

\ifdefined\showchanges
\input{sections/changelog}
\fi

\end{document}